\documentclass[useAMS,usenatbib] {mnras}
\usepackage[table]{xcolor}
\usepackage{amssymb}
\usepackage{float}

\usepackage{verbatim}
\usepackage{graphicx}
\usepackage{subcaption}
\usepackage{scalefnt}
\usepackage{amsmath}

\newcommand\T{\rule{0pt}{2.6ex}}       % Top strut
\newcommand\B{\rule[-1.2ex]{0pt}{0pt}} % Bottom strut

\usepackage[T1]{fontenc}
\usepackage{aecompl}
\usepackage{array}
\usepackage{makecell}
\newcolumntype{x}[1]{>{\centering\arraybackslash}p{#1}}
\usepackage{tikz}
\newcommand\diag[4]{%
  \multicolumn{1}{p{#2}|}{\hskip-\tabcolsep
  $\vcenter{\begin{tikzpicture}[baseline=0,anchor=south west,inner sep=#1]
  \path[use as bounding box] (0,0) rectangle (#2+2\tabcolsep,\baselineskip);
  \node[minimum width={#2+2\tabcolsep-\pgflinewidth},
        minimum  height=\baselineskip+\extrarowheight-\pgflinewidth] (box) {};
  \draw[line cap=round] (box.north west) -- (box.south east);
  \node[anchor=south west] at (box.south west) {#3};
  \node[anchor=north east] at (box.north east) {#4};
 \end{tikzpicture}}$\hskip-\tabcolsep}}

\DeclareGraphicsExtensions{.eps, .pdf}

\title[Orbital decay of BHs in clumpy galaxies]{Supermassive black hole pairs in clumpy galaxies at high redshift: delayed binary formation and concurrent mass growth}
\author[Tamburello et al.]{Valentina Tamburello,$^{1,2}$\thanks{E-mail: vtambure@physik.uzh.ch} Pedro R. Capelo,$^{1}$ Lucio Mayer,$^{1,2}$\newauthor Jillian M. Bellovary$^{3}$ and James W. Wadsley$^4$\\
$^{1}$Center for Theoretical Astrophysics and Cosmology, Institute for Computational Science, University of Zurich,\\
Winterthurerstrasse 190, CH-8057 Z{\"u}rich, Switzerland\\
$^{2}$Physik-Institut, University of Zurich, Winterthurerstrasse 190, CH-8057 Z{\"u}rich, Switzerland\\
$^{3}$Department of Astrophysics, American Museum of Natural History, Central Park West \& 79th St., New York, NY 10024, USA\\
$^{4}$Department of Physics and Astronomy, McMaster University, 1280 Main St. W, Hamilton, ON L8S 4M1, Canada}
\begin{document}
\pagerange{\pageref{firstpage}--\pageref{lastpage}} 
\maketitle
\label{firstpage}
\begin{abstract}
Massive gas-rich galaxy discs at $z \sim 1-3$ host massive star-forming clumps with typical baryonic masses in the range $10^7-10^8$~M$_{\odot}$ which can affect the orbital decay and concurrent growth of supermassive black hole (BH) pairs. Using a set of high-resolution simulations of isolated clumpy galaxies hosting a pair of unequal-mass BHs, we study the interaction between massive clumps and a BH pair at kpc scales, during the early phase of the orbital decay. We find that both the interaction with massive clumps and the heating of the cold gas layer of the disc by BH feedback tend to delay significantly the orbital decay of the secondary, which in many cases is ejected and then hovers for a whole Gyr around a separation of 1--2 kpc. In the envelope, dynamical friction is weak and there is no contribution of disc torques: these lead to the fastest decay once the orbit of the secondary BH has circularised in the disc midplane. In runs with larger eccentricities the delay is stronger, although there are some exceptions. We also show that, even in discs with very sporadic transient clump formation, a strong spiral pattern affects the decay time-scale for BHs on eccentric orbits. We conclude that, contrary to previous belief, a gas-rich background is not necessarily conducive to a fast BH decay and binary formation, which prompts more extensive investigations aimed at calibrating event-rate forecasts for ongoing and future gravitational-wave searches, such as with Pulsar Timing Arrays and the future evolved Laser Interferometer Space Antenna.
\end{abstract}
%
%%%%%%%%%%%%%%%%%%%%%%%%%%%%%
%
\section{Introduction}\label{Intro}
%
%%%%%%%%%%%%%%%%%%%%%%%%%%%%%
%
Supermassive black hole (BH) pairs at sub-kpc scales are expected to form in galactic nuclei during the hierarchical assembly of structures \citep[e.g.][]{Begelman1980,Mayer2007,Chapon2013}. In galaxy mergers, indeed, BH pairs can undergo orbital decay via dynamical friction, form a bound binary and eventually coalesce via the emission of gravitational waves (GW). The time-scales associated with the different phases of the BH pairing up to coalescence are still quite uncertain, with expectations for the overall process until coalescence ranging between $\sim10^8$~yr to more than a Hubble time \citep[e.g.][]{Mayer2013}.  Such uncertainties are problematic when one tries to infer the expected frequency of dual as well as binary active galactic nuclei (AGN), and they are of concern for the upcoming effort of future space-born GW experiments such as the evolved Laser Interferometer Space Array (eLISA), which await some robust predictions of  event rates from the theory. The formation of a bound binary is expected to occur when the BHs reach a separation of a few pc to a few tens of pc for BH masses in the range $10^6 - 10^9$~M$_{\odot}$. It is the first necessary step towards coalescence. Hydrodynamical simulations have shown that this first phase is particularly effective when there is enough gas in galactic nuclei, as dynamical friction leads to a stronger drag than in purely stellar backgrounds, leading to binary formation in only a few Myr after a major merger of two galaxies is completed \citep{Mayer2007, Chapon2013}. The same process can take $10^8-10^9$~yr in minor mergers (mass ratios 1:4 -- 1:10) as the light secondary BH is stripped of its surrounding massive baryonic core due to ram pressure and prolonged tidal effects \citep{Callegari2011, VanWassenhove2014, Capelo2015, Capelo16}.

However, recent simulations considering the clumpy multi-phase nature of the interstellar medium (ISM) in circumnuclear discs forming after the merger have shown that BHs at the low-mass end ($10^6 - 10^7$~M$_{\odot}$) can be ejected from the disc plane due to gravitational encounters and/or perturbations by a few very massive Giant Molecular Clouds (GMCs), which results in a stifling of dynamical friction and an increase of the binary formation time to $50-100$~Myr even in  major mergers \citep{Fiacconi2013, Roskar2015}. The delay is more moderate if the secondary BH is on a circular orbit as the encounter probability with perturbers is reduced \citep{DelValle2014}, but eccentric orbits are expected to result from the galaxy merger \citep[e.g.][]{Mayer2007,Roskar2015} and circularisation by dynamical friction is effective only at $\sim 1$ kpc separations and below \citep{Mayer2013}. Since the effect of such perturbations is important only when the BH has a mass of the order of that of the perturbers, for more massive BHs the gravitational perturbations of GMCs would be negligible \citep{Fiacconi2013}.

There is however a potential complication when placing the BH pairing process in the proper context of galaxy formation end evolution. Indeed, observations of the last decade have revealed that the majority of massive galaxies at high redshift, those that should host the most massive BHs, are gas-rich star-forming discs that appear clumpy on a much larger scale than that of present-day galaxies \citep{Elmegreen2005, Elmegreen2009}. Indeed there are claims of giant star-forming clumps with masses of up to $10^{10}$~M$_{\odot}$ lurking in such galaxies \citep{Tacconi2010, Zanella2015}, and many in the range $10^8-10^9$~M$_{\odot}$ have been observed \citep{Guo2012, Tacconi2013}. The origin of such giant clumps could be partly caused by in-situ fragmentation due to gravitational instability (GI) of the massive gas disc, and partly due to ex-situ contributions by minor mergers \citep{Mandelker2015}. \citet{GaborBournaud2013} have also studied the accretion on to a central BH in such a clumpy disc, arguing that it is enhanced as clumps migrate to the centre and increase the amount of gas fed to the central regions. Recently, new high-resolution hydrodynamic simulations with stronger stellar feedback tuned to reproduce galaxy stellar masses in cosmological simulations have shown that the typical mass scale of in-situ clumps formed by GI is actually in the range $10^7 - 10^8$~M$_{\odot}$, which can be explained on theoretical grounds \citep{Tamburello2015}. The larger masses often inferred could actually trace complexes of several individual $10^7$~M$_{\odot}$ clumps that are not resolved yet by typical observations \citep{Behrendt2015}, an interpretation that seems supported by the line-of-sight velocity measurements and by the smaller masses and sizes always found in lensing observations of clumpy galaxies \citep{Swinbank2009, Jones2010, Livermore2012, Livermore2015}.

Despite such recent revisions, it is almost indisputable that such massive high-redshift galaxies do possess an oversized version of GMCs in present-day galaxies, with masses that are {\it conservatively} in the range $10^7 - 10^8$~M$_{\odot}$, and can sometimes become larger, up to $10^9$~M$_{\odot}$, due to clump-clump mergers \citep{Tamburello2015}. Therefore, such clumps are massive enough to possibly interfere with the pairing of BHs having masses even exceeding $10^7$~M$_{\odot}$.

Studying such interference, if any, and its potential effect on the orbital decay of BH pairs, is the goal of this paper. Since minor rather than major mergers are the most typical kind of mergers expected during hierarchical galaxy assembly, we will study BH pairs with mass ratio 1:5 (even if in some simulations it becomes lower, 1:6, after less than $100$~Myr; see also \citealt{Callegari2009, Callegari2011}). Furthermore, as such massive clumps form from fragmentation of the large-scale galactic disc in the first place, hence at kpc scale, addressing their effect requires modelling the earliest phase of the BH pairing process, when the secondary BH has already settled in the disc of the primary galaxy, but is still far from the circumnuclear-disc region. In this paper we will thus focus on simulating the evolution of a BH pair until its separation falls below 100~pc, therefore leaving the study of binary formation and subsequent sub-pc scale decay to future work. Finally, we will also address the effect of the massive clumps on the concurrent growth of the two BHs, studying whether or not mass accretion is enhanced in a clumpy medium.

The paper is organised as follows. Section~2 describes the initial conditions and design of the simulations, Section~3 the results, taking also into account the role of the clumpy ISM on BH mass growth and the role of a stellar spheroid analytically added. Section~4, finally, summarises the paper with our conclusions.

%
%%%%%%%%%%%%%%%%%%%%%%%%%%%%%
%

\section{Simulations}\label{Section2}

%
%%%%%%%%%%%%%%%%%%%%%%%%%%%%%
%

\begin{table*}
\vspace {1 mm}
\centering
\begin{tabular}{|c|c|c|c|c|c|c|c|c|c|c|}%{|l|l|l|l|l|l|}
\hline
\bf{Model} &  \bf{vel. [km/s]}    & \bf{c}    &  $\mathbf{f_{gas}}$    &  $\mathbf{M_{vir}[M_{\odot}]}$  &  $\mathbf{M_{gas}[M_{\odot}]}$  &  $\mathbf{M_{\star}[M_{\odot}]}$  &  $\mathbf{m_{gas} [M_{\odot}]}$  &  $\mathbf{R_d [kpc]}$  &  $\mathbf{M_{BH1} [M_{\odot}]}$ \T \B \\ %&  $\mathbf{M_{BH2} [10^8 M_{\odot}]}$\\
\hline
Control  &  150  &  10  &  0.3  &  $1.3 \times 10^{12}$  & $1.35 \times 10^{10}$ &  $3.16 \times 10^{10}$ & $1.35 \times 10^5$ &  1.55  &  $2.4 \times 10^8$ \T \B \\%&  0.48\\
Clumpy &  180  &  6    &  0.5  &  $2.5 \times 10^{12}$  &  $3.9 \times 10^{10}$  &  $3.9 \times 10^{10}$   &  $3.9 \times 10^5$   &  2.27  &  $4.9 \times 10^8$ \B \\%&  0.98\\
\hline
\end{tabular}
\caption{The table shows the main features of our two models, the control one (first row) and the clumpy one (second row). Column 2: velocity at virial radius; Column 3: concentration; Column 4: gas fraction; Column 5: virial mass; Column 6: gas mass; Column 7: stellar mass; Column 8: gas particle mass; Column 9: disc scale length; Column 10: mass of the primary BH (the mass of the secondary BH is initially 1/5 of $M_{\rm BH1}$)}
\label{table1}
\end{table*}

We perform a set of numerical simulations of isolated galaxies to study the evolution of BH pairs in clumpy discs at high redshift. Simulations are performed using the $N$-body+smoothed-particle-hydrodynamics (SPH) code {\scshape gasoline2}, which uses a pressure-energy formulation of the hydrodynamical force, thermal and metal diffusion, and a Wendland kernel to remove artificial surface tension and resolve two-fluid instabilities (see e.g. \citealt{Keller2015, Governato2015, Tamburello2015}).

Since we wish to study the dynamics of BH pairs in clumpy high-redshift discs, we use two models from our previous work, in particular Models~7 and 11, listed in Table~1 of \citet{Tamburello2015}. 
Model~11, the most massive one with concentration 6, high gas fraction as typical for $z \sim 2$ (50 per cent of the disc mass) and a very massive disc ($8 \times 10^{10}$~M$_{\odot}$), is at the high-mass end of the observed galaxies at $z \sim 2$, having $V_{max} \sim 350$~km~s$^{-1}$ \citep{Wisnioski2015}, and does produce several massive clumps, of mass up to $10^9$~M$_{\odot}$, in our previous simulations without BHs \citep{Tamburello2015}. The other one, Model~7, with concentration 10, gas fraction 0.3 and disc mass $4 \times 10^{10}$~M$_{\odot}$, undergoes only transient fragmentation into a few clumps and stabilises into a configuration with prominent spiral patterns after $\sim 200$~Myr. Model~7 has a mass and maximum circular velocity ($\sim 250$~km~s$^{-1}$) that are close to the average expected for galaxies at $z > 1$, hence its behaviour should be regarded as more representative of the typical disc galaxy at those high redshifts. Since it undergoes only some transient fragmentation at the beginning, the BHs will evolve in a considerably smoother disc, hence we can consider Model~7 as the {\it control run} for any clump-related dynamical effect found with Model~11.

Hereafter we will refer to Model~11 as the {\it clumpy} model and to Model~7 as the {\it control} model, but we should bear in mind that Model~7 is a marginally unstable disc and not a completely smooth disc as those considered in previous works on minor mergers \citep[e.g.][]{Callegari2009,Callegari2011}. We caution the reader that to understand the effect of clumps alone is not straightforward, since we also change other variables, such as mass and gas fraction, but we address it in our comparison. One solution would have been to use different equations of state for the same system. However, we believe that considering two systems with the same physical parameters, but different cooling prescriptions, would make the study less applicable to the actual target, namely to reproduce conditions actually expected in $z \gtrsim 1$ galaxies \citep{Tamburello2015}.

Both models are described in Table~1 of \citet{Tamburello2015}. They were built as three-component `dark-matter halo+stellar disc+gaseous disc' models using the technique originally developed by \citet{Hernquist1993} and relaxed adiabatically for 1~Gyr to avoid sudden amplification of particle noise by the high self-gravity of the disc. The modelling choices are thoroughly described in \citet{Tamburello2015}, to which the reader should refer. After 1~Gyr the BHs are inserted as massive collisionless particles and at the same time radiative cooling is switched on. In all runs we adopt metal-dependent cooling using the implementation of CLOUDY \citep{Ferland_et_al_1998} tables in the hydrodynamical code as described by \citet{Shen2010}.

For the sub-grid prescriptions, we adopt the recipes described in \citet{Stinson2006} for star formation and (blastwave) supernova feedback: in particular, a star can form if the progenitor gas particle has a temperature lower than $3 \times 10^4$~K and is denser than $10$~cm$^{-3}$, with a star formation efficiency $\epsilon_{\rm SF} = 0.01$, while supernovae release an energy of $4 \times 10^{50}$~erg. In some of the runs (see Table~\ref{table2}) we also add accretion on to the BHs and thermal BH feedback as described in \citet{Bellovary2010} and \citet{Bonoli2015}. Basically BHs are assumed to accrete gas isotropically, according to a Bondi-Hoyle-Lyttleton \citep[hereafter, Bondi;][]{Bondi1952} accretion recipe:

\begin{equation}
\dot{M}_{\rm acc} = \frac{4\pi \alpha G^2 M^2_{\rm BH}\rho_{\rm g}}{(c_{\rm s}^2 + v^2)^{3/2}},
\label{eq:Bondi}
\end{equation}

\noindent where $\rho_{\rm g}$ is the local gas density, $c_{\rm s}$ is the sound speed, $v$ is the relative velocity of the BH compared to the gas velocity, $G$ is the gravitational constant, $M_{\rm BH}$ is the BH mass, and $\alpha = 1$ is the BH accretion boost factor. Accretion rate is limited by the Eddington rate, which follows the equation:

\begin{equation}
\dot{M}_{\rm Edd} = \frac{4\pi G^2 M_{\rm BH}m_{\rm p}}{\epsilon_{\rm r} \sigma_{\rm T} c},
\label{eq:Eddington}
\end{equation}

\noindent where $m_{\rm p}$ is the proton mass, $\epsilon_r$ is the radiative efficiency of the accreted gas (here $0.1$), $\sigma_{\rm T}$ is the Thomson cross section, and $c$ is the speed of light in vacuum.

BH feedback, finally, follows a thermal model in which a fraction ($\epsilon_{\rm f} = 0.001$) of the mass-energy of the accreted gas is converted to radiative energy and then is isotropically distributed to the kernel of the SPH particle (without using a blastwave model to delay the cooling as in \citealt{Bellovary2010}). As we already said, at the end of the adiabatic relaxation (i.e. after 1~Gyr), we add two BHs in both models. The most massive (the primary) BH is at the centre of mass of the galaxy, while the least massive (secondary) is at an initial separation $a_0$ from the centre equal to the disc scale length radius (1.55~kpc for the control model and 2.27~kpc for the clumpy model), and has a velocity $v_0$ equal to that of the local baryonic centre of mass. The assumption on the initial separation is consistent with the typical distance from the centre that the secondary BH has following an unequal-mass galaxy merger with typical orbital parameters \citep{Callegari2011,Capelo2015}, once most of the baryonic core of the galaxy around the BH has been stripped, and hence the BH can be treated as ``naked'' as we assumed here.

The mass of the primary BH is computed using the equation in \citet{Bennert2010}:

\begin{equation}
\log M_{\rm BH} - 8 = \alpha^{\prime}(\log M_* - 10) + \beta \log (1+z) + \gamma + \sigma,
\label{eq:Bennert}
\end{equation}

\noindent where $\alpha^{\prime}$ is the slope of the relation at $z = 0$, $\beta$ describes the evolution of the scaling relation ($\beta = 0$ means no evolution), $\gamma$ is the intercept of the relation at $z = 0$, and $\sigma$ is the intrinsic scatter. For these parameters, we use the values given in Table~3 of \citet{Bennert2010}, obtained fitting objects from \citet{Bennert2010} and \citet{Merloni2010}: $\alpha^{\prime} = 1.12$, $\beta = 1.15$, $\gamma = -0.68$, and $\sigma = 0.16$. The initial mass of the secondary BH is 1/5 of that of the primary one. The BHs mass is reported in Table~\ref{table1}.

\begin{table*}
\vspace {1 mm}
\centering
\begin{tabular}{|c|c|c|c|c|c|c|c|c|c|cl}%{|l|l|l|l|l|l} 
\hline
Model & vel. [km~s$^{-1}$] & $c$    	& $f_{\rm gas}$ & $f$  	& $e_0$  	& $\alpha$    	& $\epsilon_{\rm f}$ & $\epsilon_{\rm r}$ & $t_{\rm cp}$ [Gyr] \T \B \\
\hline
Control-f0-noAcc 		& 150	& 10	& 0.3		& 0		& 0		& 0	& -   		&  -  		& 0.169 	\T \B \\
Control-f0-stdAcc-stdFB 			& 150    	& 10 	& 0.3  	& 0  		& 0     	& 1  	& 0.001   	& 0.1    	& 0.378 	\B 	\\
Control-f0-stdAcc-lowFB 	& 150    	& 10 	& 0.3  	& 0  		& 0     	& 1  	& 0.0001 	& 0.1    	& 0.276 	\B 	\\
Control-f0-stdAcc-verylowFB 	& 150    	& 10 & 0.3  	& 0  		& 0     	& 1  	& 0.00001	& 0.1    	& 0.158	\B 	\\
Control-f0-highAcc-stdFB 	& 150    	& 10 & 0.3  	& 0  		& 0     	& 10	& 0.001  	& 0.1    	& 0.423 	\B 	\\
Control-f02-noAcc 		& 150    	& 10 & 0.3  	& 0.2  	& 0.2  	& 0	& -  		& -  		& 0.240 	\B 	\\
Control-f02-stdAcc-stdFB 		& 150    	& 10 & 0.3  	& 0.2  	& 0.2  	& 1  	& 0.001  	& 0.1  	& 0.488 	\B 	\\
Control-f1-noAcc 		& 150    	& 10 & 0.3  	& 1  		& 0.7  	& 0  	& -  		& -  		& 0.292 	\B 	\\
Control-f1-stdAcc-stdFB 			& 150    	& 10 & 0.3  	& 1  		& 0.7  	& 1  	& 0.001  	& 0.1  	& -	\B 	\\
Control-f2-noAcc 		& 150    	& 10 & 0.3  	& 2  		& 0.9  	& 0 	& -  		& -   		& 0.374 	\B 	\\
Control-f2-stdAcc-stdFB 			& 150    	& 10 & 0.3  	& 2  		& 0.9  	& 1  	& 0.001    	& 0.1    	& 0.699 	\B 	\\
\hline
Clumpy-f0-noAcc 		& 180    	& 6	& 0.5 	& 0  		& 0  		& 0  	& -    	& -   		& 0.146 	\T \B \\
Clumpy-f0-stdAcc-stdFB 			& 180    	& 6 	& 0.5    	& 0  		& 0  		& 1  	& 0.001   	& 0.1    	& - 		\B	\\
Clumpy-f0-stdAcc-lowFB 	& 180    	& 6 	& 0.5    	& 0  		& 0  		& 1  	& 0.0001 	& 0.1    	& - 		\B 	\\
Clumpy-f0-stdAcc-verylowFB 	& 180    	& 6 	& 0.5    	& 0  		& 0  		& 1  	& 0.00001	& 0.1    	& 0.111 	\B 	\\
Clumpy-f0-highAcc-stdFB 	& 180    	& 6 	& 0.5    	& 0  		& 0  		& 10 	& 0.001   	& 0.1    	& - 		\B 	\\
Clumpy-f02-noAcc 		& 180    	& 6 	& 0.5  	& 0.2  	& 0.2  	& 0  	& -  		& -   		& 0.166 	\B 	\\
Clumpy-f02-stdAcc-stdFB 		& 180    	& 6 	& 0.5    	& 0.2  	& 0.2  	& 1  	& 0.001   	& 0.1    	& 0.792 		\B 	\\
Clumpy-f1-noAcc 		& 180    	& 6 	& 0.5  	& 1  		& 0.7  	& 0  	& -  		& -   		& - 		\B 	\\
Clumpy-f1-stdAcc-stdFB 			& 180    	& 6 	& 0.5    	& 1  		& 0.7  	& 1  	& 0.001  	& 0.1    	& 0.677 	\B 	\\
Clumpy-f2-noAcc 		& 180    	& 6 	& 0.5  	& 2  		& 0.9  	& 0  	& -  		& -   		& - 		\B 	\\
Clumpy-f2-stdAcc-stdFB 			& 180    	& 6 	& 0.5    	& 2  		& 0.9  	& 1  	& 0.001   	& 0.1    	& 0.414 		\B 	\\
\hline
\end{tabular}
\caption{Scheme of the simulations. The names in Column 1 are self-explanatory: $f$ is a proxy for the eccentricity; noAcc stands for no BH accretion; stdAcc and highAcc stand for standard ($\alpha=1$) and high ($\alpha=10$) BH accretion, respectively; stdFB, lowFB, and verylowFB stand for standard ($\epsilon_{\rm f}=0.001$), low ($\epsilon_{\rm f}=0.0001$), and very low ($\epsilon_{\rm f}=0.00001$) BH feedback, respectively. Columns 2--4 (same as in Table~\ref{table1}): virial velocity, concentration, and gas fraction of each model, for easier comparisons. Columns 5--6: eccentricity [see Equation~\eqref{eq:eccentricity}]. Columns 7--9  show how we change the BH accretion and feedback parameters. Column 10: BH close-pairing (i.e. BH separation < 100~pc) time in Gyr. When no close pairing occurs within 1~Gyr, there is simply a -.}
\label{table2}
\end{table*}

In our simulations, summarised in Table~\ref{table2}, we change the eccentricity of the orbit of the secondary BH, see Table~\ref{table2}. As in \citet{Fiacconi2013}, the ratio $f$ between the radial and azimuthal components of the initial velocity $v_0$ specifies the orbit of the secondary BH. The initial velocity is $\left|{v_0}\right| = V_c(a_0)$, where $V_c(a_0)$ is the circular velocity at the initial separation. The initial eccentricity $e_0$ of the orbit is also determined by the ratio $f$ (assuming a bound Keplerian orbit, i.e. elliptical):

\begin{equation}
e_0 \simeq \sqrt{1 - \frac{1}{1+f^2}}.
\label{eq:eccentricity}
\end{equation}

Note that the BHs do start at the same distance in all runs. As a result, higher-eccentricity orbits result in smaller-pericentre distances. The number of gas particles is $10^5$, as is the number of star particles, whereas the dark matter particles are $1.2 \times 10^6$. The gravitational softening is $100$ pc for all the components. As described in \citet{Tamburello2015}, owing to the relatively high number of particles, the hydrodynamical resolution is much higher, as smoothing lengths are a factor of 10 lower than the gravitational softening in the high-density regions of the disc. We ran 22 simulations (11 per model; see Table~\ref{table2}), varying the initial eccentricity of the orbit of the secondary BH and the BH accretion and feedback parameters ($\alpha$ and $\epsilon_{\rm f}$), up to the point the BHs form a close pair (i.e. their separation falls below the gravitational softening, 100~pc). If a close pair does not form after 1~Gyr, we stop the simulation, since our simulations are isolated.

%
%%%%%%%%%%%%%%%%%%%%%%%%%%%%%
%

\section{Results}

%
%%%%%%%%%%%%%%%%%%%%%%%%%%%%%
%

\subsection{Orbital Decay}

%
%%%%%%%%%%%%%%%%%%%%%%%%%%%%%
%

As already found in previous work \citep{Callegari2011, Fiacconi2013, Roskar2015}, the BH orbital decay is affected by the presence of dense structures and clumps. The most common trend in our simulations is a delay.

\subsubsection{Control model}

\begin{figure}
\flushleft
\includegraphics[width=0.49\textwidth]{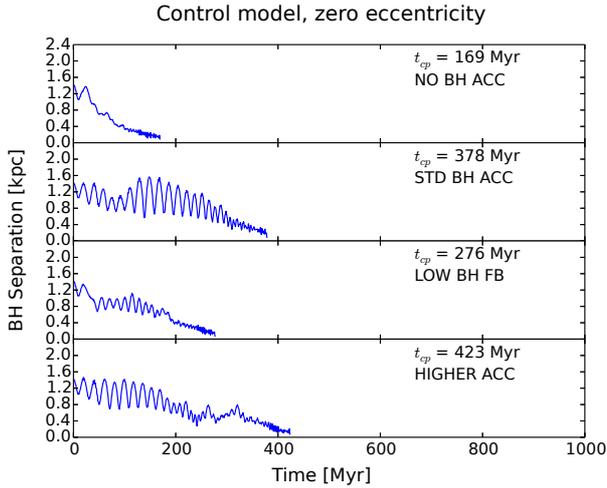}
\caption{Time evolution of the BH separation for the control model in the cases with zero eccentricity. From top to bottom: in the first panel we show the case without BH accretion; the orbital decay is fast, disturbed only by the presence of a few small clumps (before 100~Myr) that are responsible for the peak at $\sim 20$~Myr. Switching on BH accretion and feedback (from the second to the fourth panel) makes the orbital decay time-scale larger, because BH feedback rarefies the gas and dynamical friction becomes less efficient. In the third panel, $\epsilon_{\rm f}$ is reduced by a factor 10 (from 0.001 to 0.0001) compared to the second panel: reducing the BH feedback efficiency leads back to a shorter orbital decay, because the BH is not able to rarefy the gas. Finally, in the fourth panel, we increase the BH accretion boost factor, $\alpha$, from 1 to 10 [see Equation~\eqref{eq:Bondi}] and the orbital decay time-scale becomes slightly longer than in the standard case (second panel), due to the slightly higher feedback.}
\label{fig:separation_control_f0}
\end{figure}

\begin{figure*}
\flushleft
\includegraphics[width=0.49\textwidth]{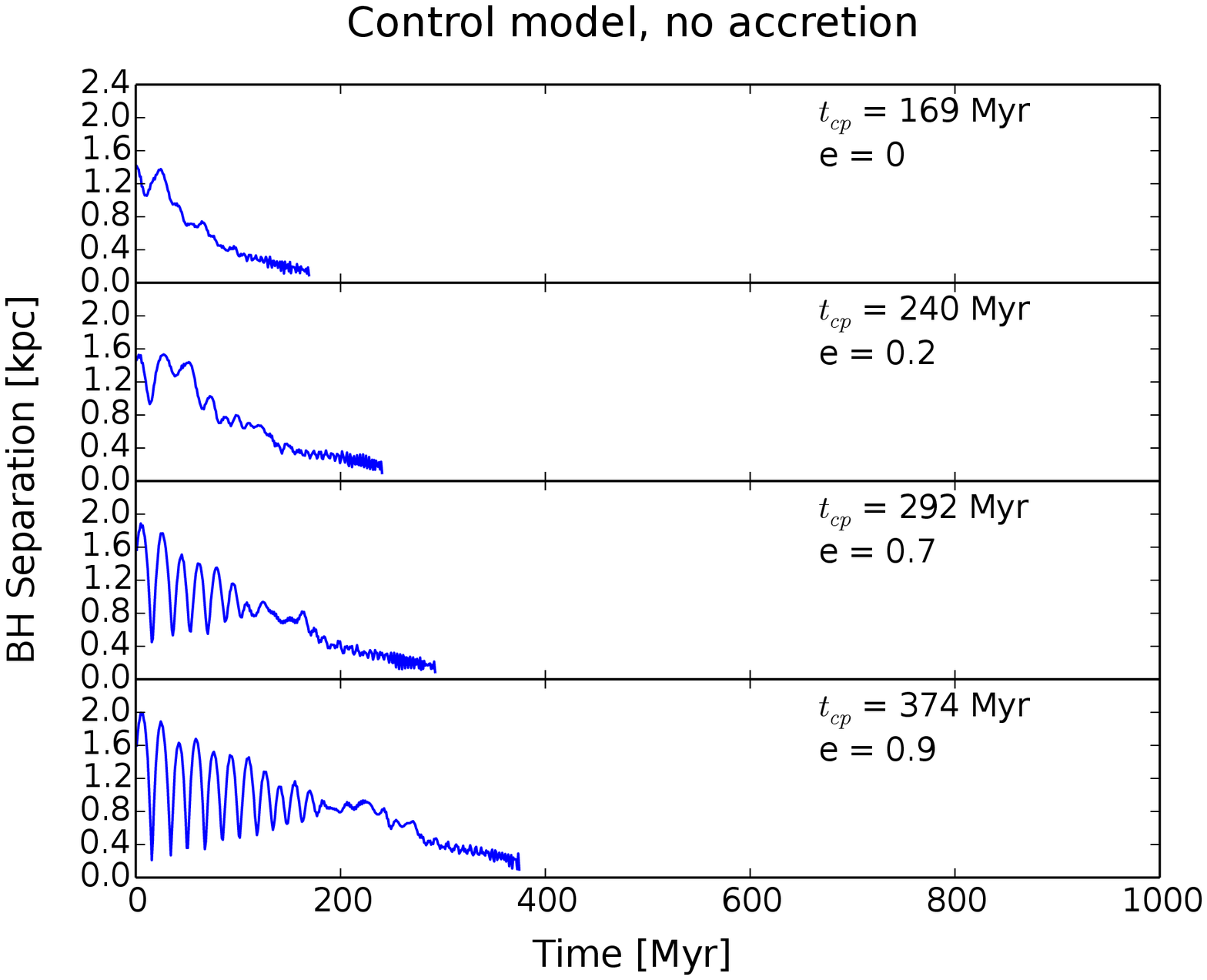}
\includegraphics[width=0.49\textwidth]{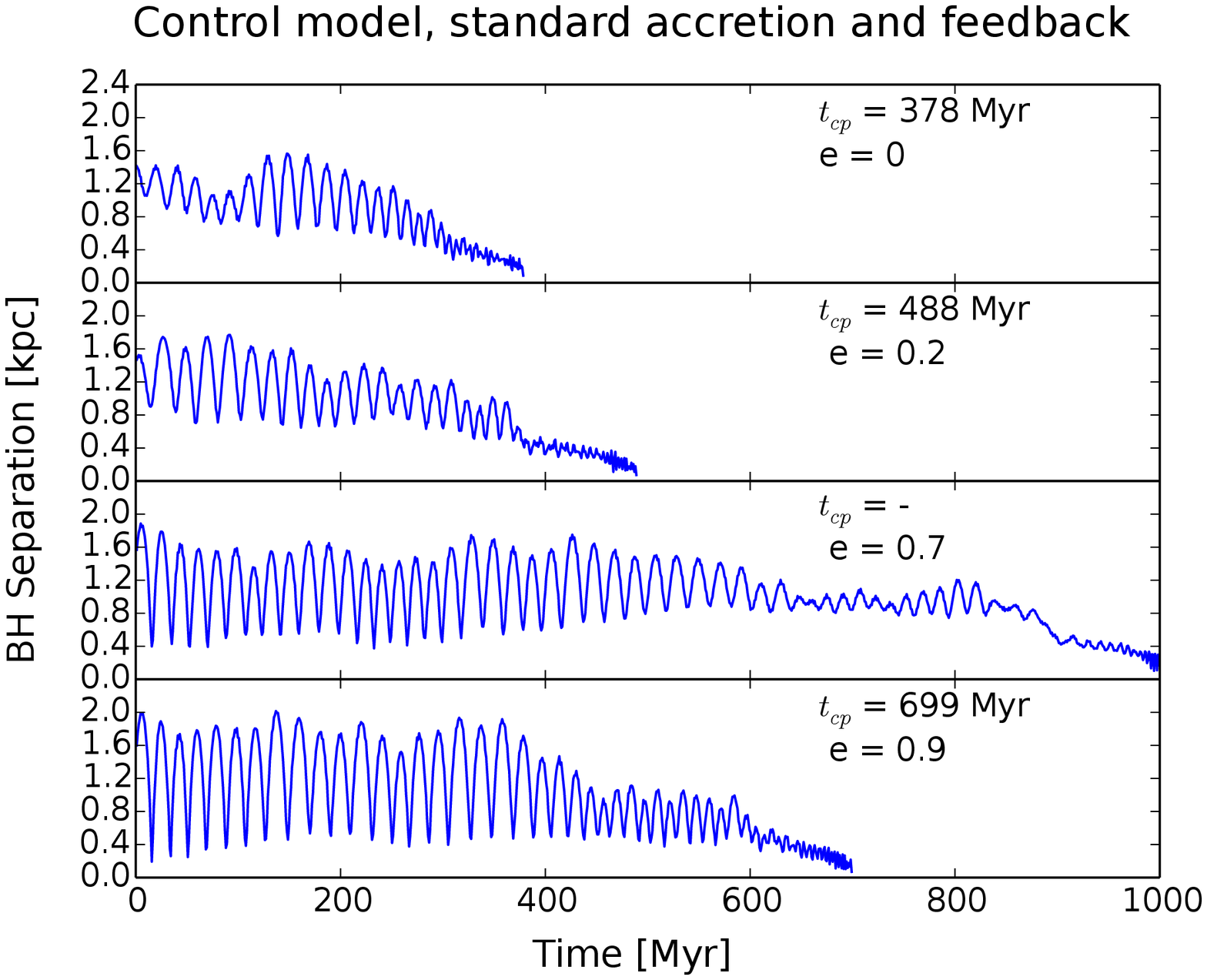}
\caption{Time evolution of the BH separation for the control model with different eccentricities: on the left we show the cases without BH accretion, whereas on the right we include BH accretion and feedback. In both cases increasing the eccentricity (from top to bottom, the eccentricity is $0$, $0.2$, $0.8$, and $0.9$) delays the orbital decay time-scale compared to the case with $e_0 = 0$. This happens, as explained in the text, because the orbit needs to circularise before, then the spiral-wave torque acts quickly and leads to a fast decay. Notice that the orbital decay in the right panels is longer than that in the corresponding left panels due to the presence of BH feedback. In the third case of the right panels we do not form a close BH pair in less than 1~Gyr (our time limit for all the simulations), but the BH separation is less than 300~pc.}
\label{fig:separation_control_diffEcc}
\end{figure*}

The control model resembles a galaxy at $z \sim 1$, as already shown in \citet{Tamburello2015}, and produces fewer clumps compared to the clumpy model, for this reason offering a simple benchmark for comparison purposes. This model has a disc mass of $4.5 \times 10^{10}$~M$_{\odot}$ and gas fraction $0.3$. The mass of the primary BH is $2.4 \times 10^8$~M$_{\odot}$, whereas the secondary BH has a mass of $4.8 \times 10^7$~M$_{\odot}$, comparable to the mass of our clumps. In the case without BH accretion and feedback, with eccentricity $e_0 = 0$, the BH orbital decay is quite smooth (see Figure~\ref{fig:separation_control_f0}, top panel) with peaks due to clumps' presence at 20--50~Myr after the beginning of the simulation. Notice that clumps form earlier with respect to the simulations in our previous work (\citealt{Tamburello2015}), probably because of the perturbing influence of the secondary BH, in analogy to the triggered fragmentation phenomenon well documented for self-gravitating protoplanetary discs \citep{Meru2015}.

\begin{figure}
\flushleft
\includegraphics[width=0.49\textwidth]{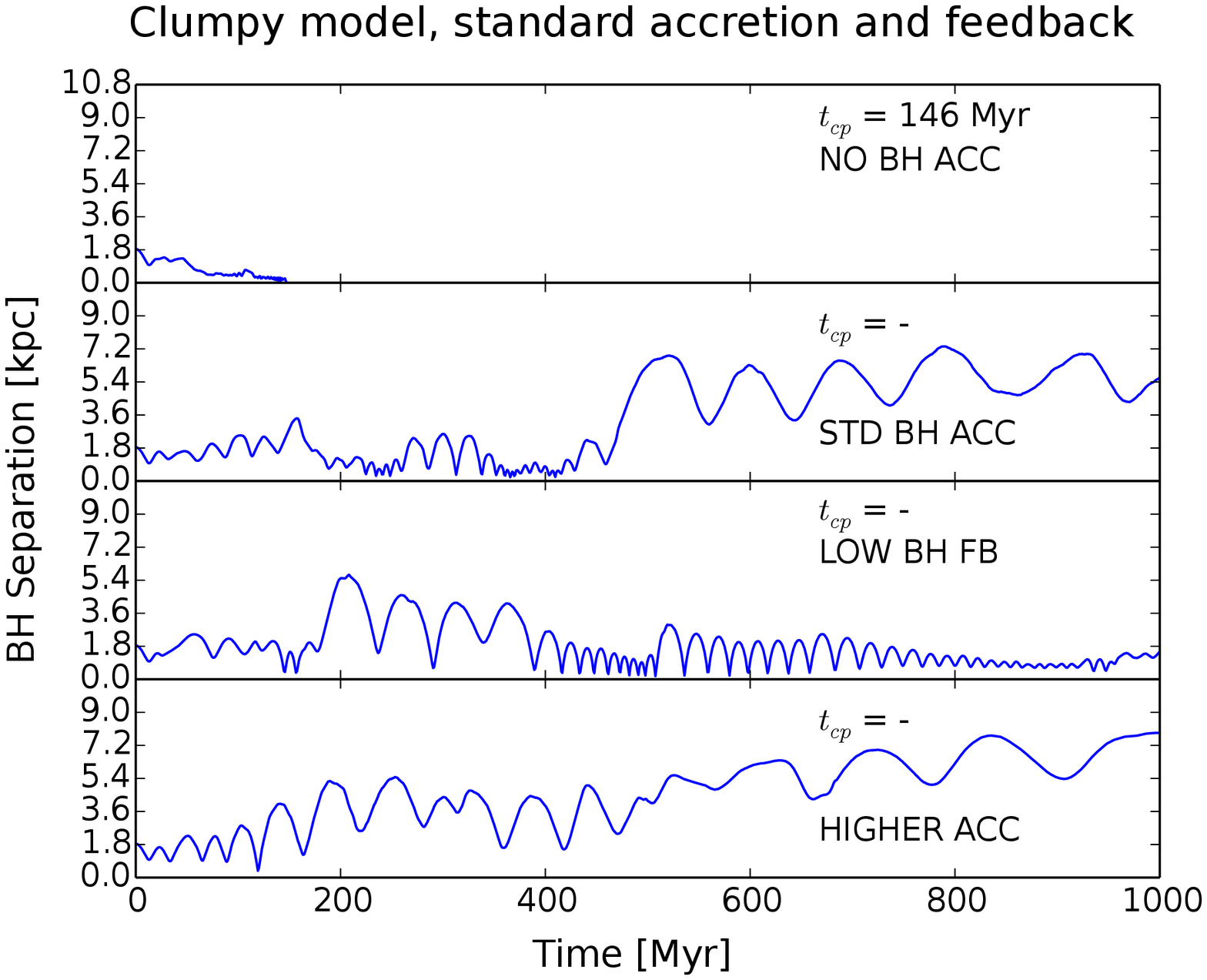}
\caption{Time evolution of the BH separation for the clumpy model in the cases with zero eccentricity. From top to bottom: in the first panel we show the case without BH accretion. Switching on BH accretion and feedback (from the second to the fourth panel) makes the orbital decay time-scale larger and the BH close pairing never happens (within 1~Gyr). In the third panel, $\epsilon_{\rm f}$ is reduced by a factor 10 (from 0.001 to 0.0001) compared to the second panel, but the presence of many clumps does not allow the BH pair formation. Finally, in the last panel we increase the BH accretion boost factor, $\alpha$, from 1 to 10 [see Equation~\eqref{eq:Bondi}] and the close BH pair never happens.}
\label{fig:separation_massive_f0}
\end{figure}

\begin{figure*}
\flushleft
\includegraphics[width=0.49\textwidth]{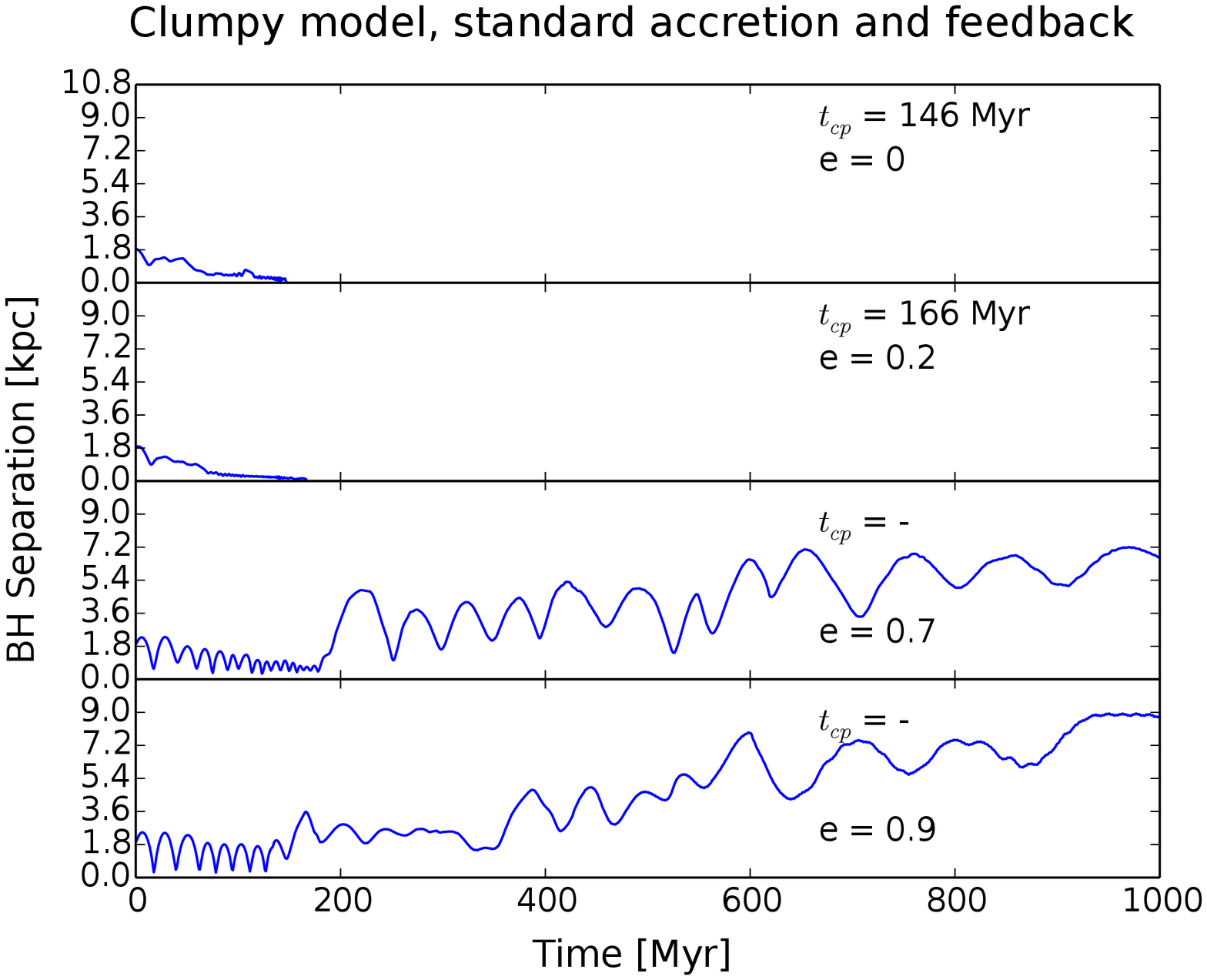}
\includegraphics[width=0.49\textwidth]{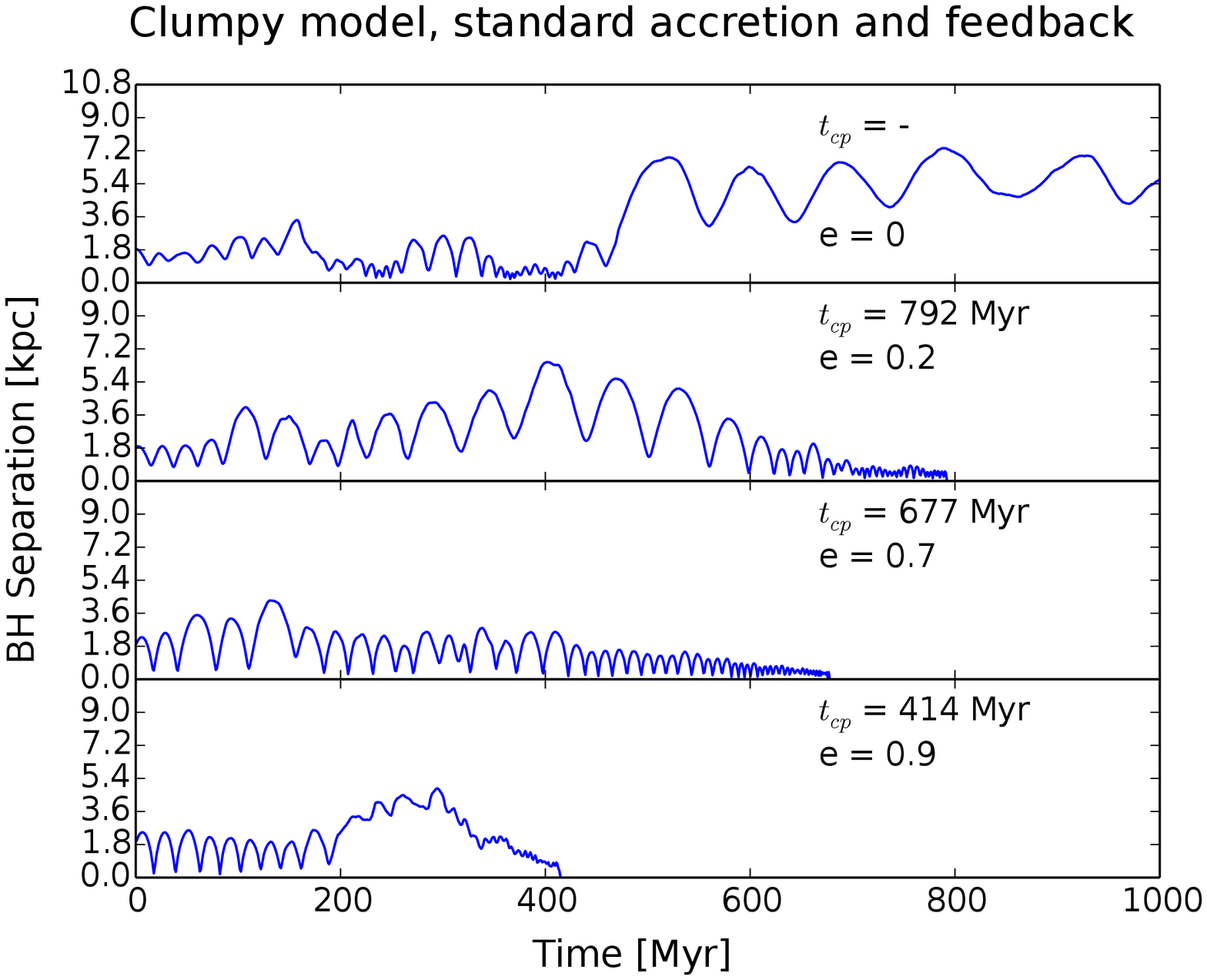}
\caption{Time evolution of the BH separation for the clumpy model with different eccentricities: without BH accretion and feedback on the left, and with BH accretion and feedback on the right. Eccentricity increases from top to bottom ($0$, $0.2$, $0.8$, and $0.9$). In the clumpy cases the situation is more complicated than in the control cases. In the cases without BH accretion and low eccentricity, $0$ and $0.2$ (first and second panels on the left side) the orbital decay happens faster than in the control model (see Figure~\ref{fig:separation_control_diffEcc}, top left panels, for a comparison), while in the other cases the pairing never happens, because a high eccentricity favours clump encounters that can scatter the secondary BH. Adding BH accretion and feedback (right side), rarefies and heats the gas, making less efficient the dynamical friction and therefore causing a delay, but encounters with clumps also change continuously the secondary BH dynamics.}
\label{fig:separation_clumpy}
\end{figure*}

The clumps are still short-lived as in our original runs, hence they disturb only marginally the orbital decay, with the BH close-pairing time (i.e. BH separation < 100~pc) happening in less than 200~Myr. As we turn on BH accretion, we expect a faster orbital decay, since the secondary BH can increase its mass and so both dynamical friction and disc torques should be more efficient. This, however, does not happen because the secondary BH does not grow significantly, as discussed further below, and because of the complex interplay with BH feedback, which rarefies the gas, making the orbital decay time-scale longer: the density of the gas around the secondary BH is lower in the case with accretion than in the case with no accretion, by a factor of $\sim$10 on average, especially in the early times. Because the dynamical friction time-scale scales with the dynamical time, which is proportional to the inverse of the square root of the density, a factor of $\sim$10 in density implies a factor of $\sim$3 in BH pairing time-scale, which is close to what seen in Figure~\ref{fig:separation_control_f0} (compare the top two panels: the BH close-pairing time increases from 169 to 378~Myr). If we try to decouple the effect of BH accretion from that of BH feedback, to understand their role, we notice that when we decrease the BH feedback efficiency (from $\epsilon_{\rm f} = 0.001$ to 0.0001; third panel of Figure~\ref{fig:separation_control_f0}), the BH close-pairing time becomes again short (276~Myr), as expected. We double-checked this assumption by reducing the BH feedback efficiency even further (see Table~2; $\epsilon_{\rm f} = 0.00001$, not shown in Figure~\ref{fig:separation_control_f0}), obtaining a BH close-pairing time of 158~Myr, comparable to the case without accretion, showing that the effect of {\it only} BH accretion is negligible and that BH feedback is responsible for the change in the BH orbital time-scale. We also tested a case where we increased the value of the BH accretion boost factor $\alpha$ [from 1 to 10; see Equation~\eqref{eq:Bondi}]. Even if one could in principle expect an increase in $\dot{M}_{\rm acc}$ by a factor of $10$, this does not happen because of self-regulation, and the increase is very small. As a consequence, the increase in BH close-pairing time is also negligible (423 versus 378~Myr; see the second and fourth panels in Figure~\ref{fig:separation_control_f0}).

Of course, also eccentricity plays an important role in the orbital decay. In particular, in the absence of BH accretion, the higher the eccentricity, the higher the time needed to form a close pair (see Figure~\ref{fig:separation_control_diffEcc}, left panels), consistent with what found in \citet{Fiacconi2013}. In the presence of BH accretion, the behaviour is similar, even if more complex, because feedback makes the gas warmer and spiral waves become weaker; we still observe an increase of the orbital decay time-scale increasing the eccentricity. One could think that, due to high eccentricity, the secondary BH should form a close pair earlier since it approaches more the denser region around the primary BH, but this does not happen. As discussed in \citet{Mayer2013} and in \citet{Fiacconi2013}, the fastest phase of the decay is not due to dynamical friction, but to the spiral wave torques (like in planet migration), because it is only then that the angular momentum decreases fast. Basically, at the beginning, the orbit needs to circularise, because when the orbit is eccentric the spiral wave torque is weak, since it is changing sign along the orbit. Once circularisation is complete, the spiral wave torque acts quickly and leads to a fast decay. This is clear in Figure~\ref{fig:separation_control_diffEcc}, both left and right panels: orbits circularise first (see the pericentres and apocentres getting more similar), then the decay happens fast. Moreover, comparing the first panel with the last one on the left of Figure~\ref{fig:separation_control_diffEcc}, one can notice that, once the circularisation happens, the close pair formation happens in the same time, i. e. $\sim 200$~Myr or less.

%
%%%%%%%%%%%%%%%%%%%%%%%%%%%%%
%

\subsubsection{Clumpy model}

%
%%%%%%%%%%%%%%%%%%%%%%%%%%%%%
%

The clumpy model, with a higher gas fraction (0.5) and disc mass ($8 \times 10^{10}$~M$_{\odot}$) than those in the control model, resembles a galaxy at $z \sim 2$ even if, as discussed in our previous work \citep{Tamburello2015}, it is an extreme case for high-redshift galaxies \citep[see][]{Wisnioski2015}. The mass of the primary BH is $4.9 \times 10^8 $~M$_{\odot}$, whereas the secondary BH has a mass of $9.8 \times 10^7$~M$_{\odot}$. This case gives rise to a markedly different nature of the gaseous background in which the secondary BH evolves, since the massive self-gravitating disc fragments into many more, more massive, and longer-lived clumps than in the control model \citep[see][]{Tamburello2015}. The results are not as straightforward as in the control model, as the multiple massive clumps and strongly perturbed disc background with a vigorous spiral pattern (in part forced by the clumps) lead to a stochastic behaviour of the orbital dynamics of the secondary BH.

As in \citet{Roskar2015}, the orbital decay is strongly affected by the complex gravitational interactions between the secondary BH and individual clumps. Depending on how the interaction takes place, a clump encounter can indeed accelerate {\it or} delay the decay, as it can exert both positive and negative torques. Furthermore, overdense spiral arms also exert strong non-linear torques, which can also be either positive o negative, a situation reminiscent of the migration of gas giant planets in self-gravitating protoplanetary discs \citep{Baruteau_et_al_2011,Malik_et_al_2015}. BH accretion and feedback complicate further the scenario by affecting the density and temperature of the background. By looking at the overall ensemble of clumpy-disc simulations, it is evident that the most common outcome is a substantial delay of the orbital decay (see Figures~\ref{fig:separation_massive_f0} and \ref{fig:separation_clumpy}). In particular, this delay always occurs with BH accretion and feedback (Figure~\ref{fig:separation_massive_f0}), which confirms the trend seen also in the control run when BH accretion and feedback are considered. Indeed, there are runs where the orbit of the secondary BH does not decay in even 1~Gyr (and, being isolated simulations, we decided  to interrupt such runs as they would not be meaningful on longer time-scales; see e.g. \citealt{Tamburello2015} for a discussion).

\begin{figure*}
\includegraphics[width=0.49\textwidth]{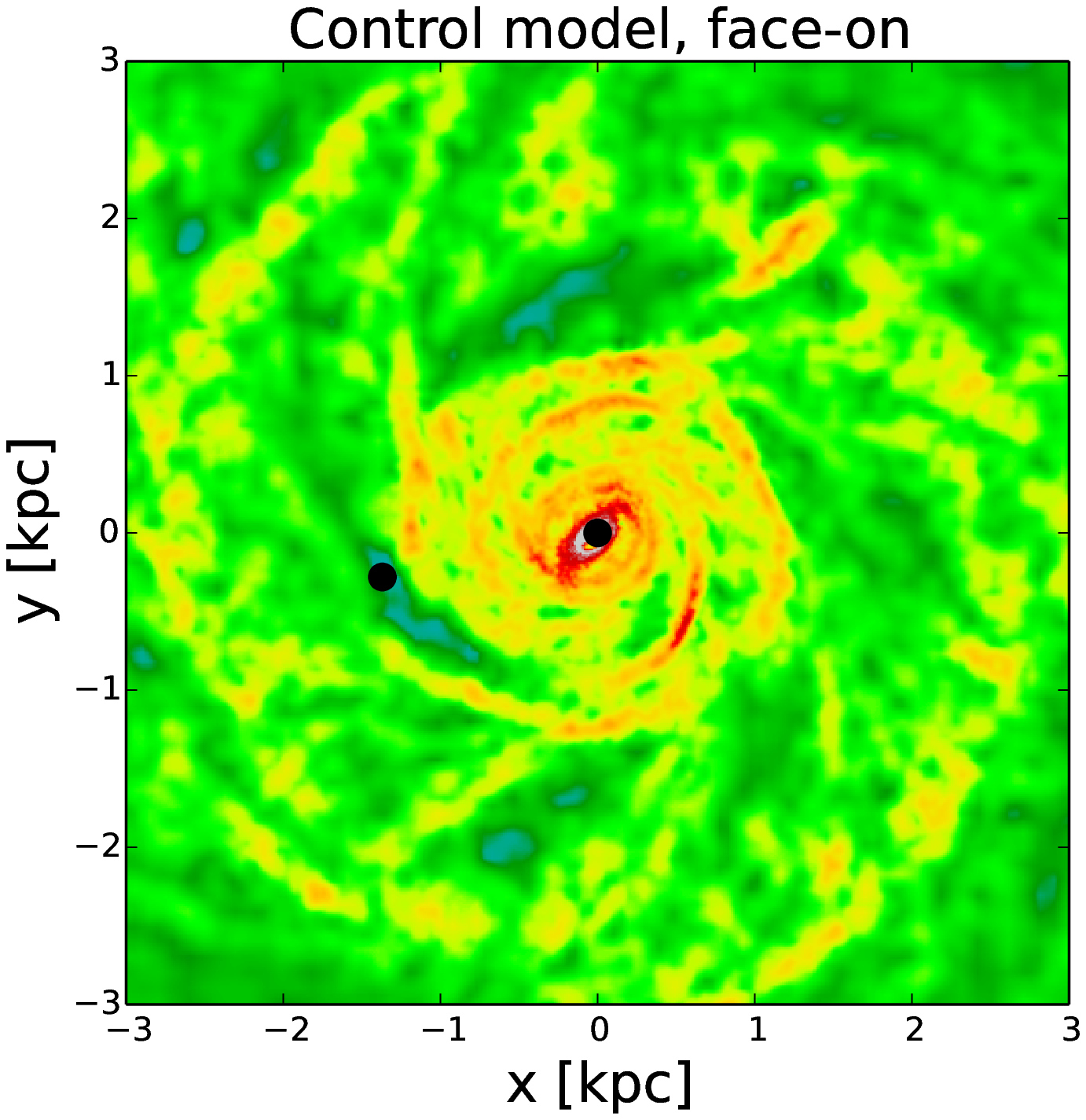}
\includegraphics[width=0.49\textwidth]{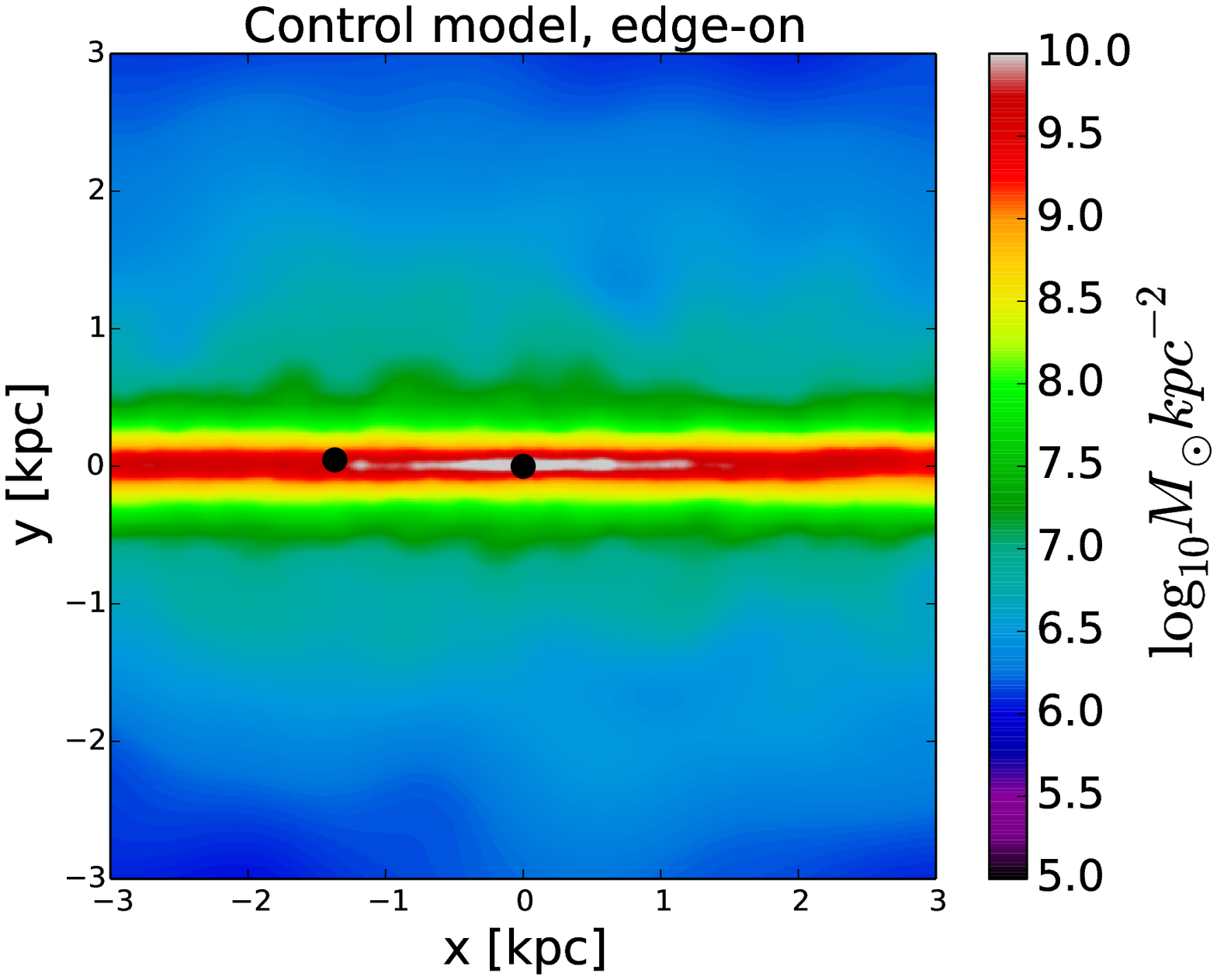}\\
\includegraphics[width=0.49\textwidth]{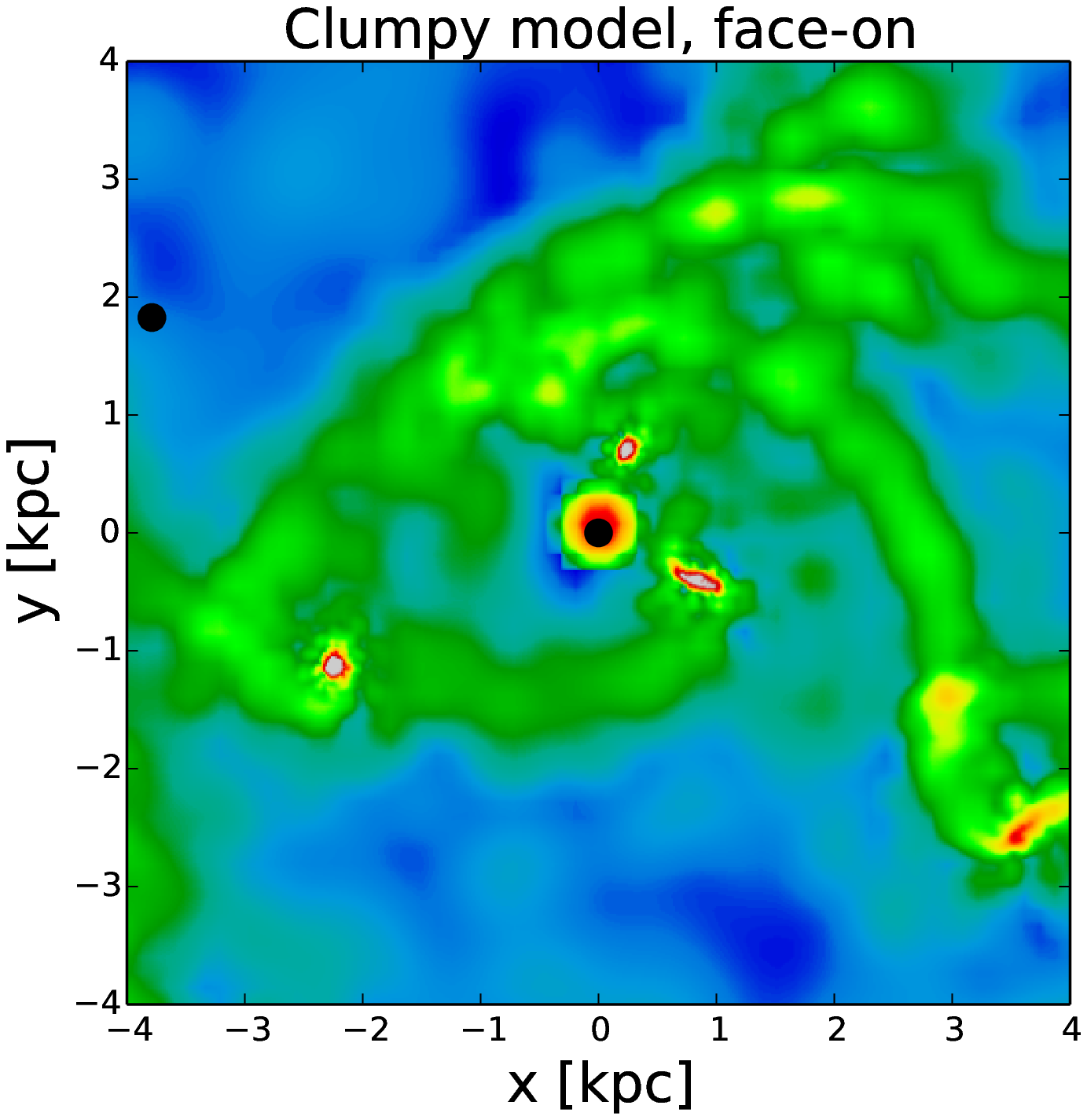}
\includegraphics[width=0.49\textwidth]{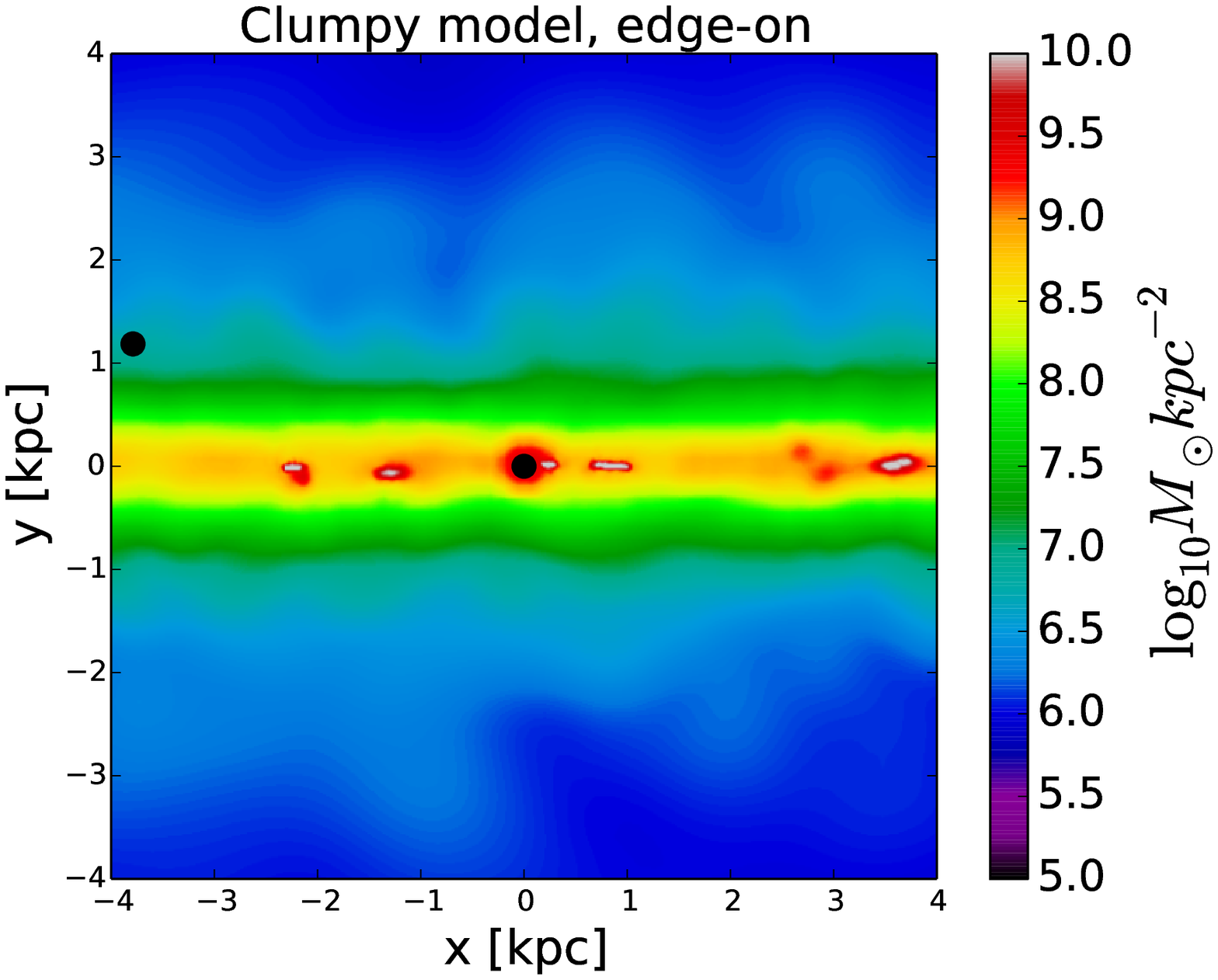}
\caption{Surface density maps, face-on and edge-on, for the control model (top panels) at 42~Myr and the clumpy model (bottom panels) at 345~Myr, with BH accretion, feedback, zero eccentricity and $\alpha = 10$. Times are chosen to show the maximum separation of the secondary BH from the midplane in the first 400~Myr. In the control model, the secondary BH is always close to the midplane  (the maximum distance in the $z$-axis is much lower than $\sim 500$~pc), because the clumps are small, disappear in the first $\sim 100$~Myr and do not affect BH dynamics. In the clumpy model, the secondary BH can exceed 1~kpc from the midplane and reach hotter and less dense regions due to clump interactions.}
\label{fig:low_density_region}
\end{figure*}

The main reason for the frequent occurrence of a delayed orbital decay in clumpy discs is that the clump-BH interaction results in perturbations of the orbit of the secondary BH. These perturbations can either increase the radius of the orbit of the secondary BH within the disc, or lead to vertical oscilations up to $\sim 2$~kpc above the disc midplane: at this distance the density is so low that dynamical friction and disc torques become very inefficient (see Figure~\ref{fig:low_density_region}). This is a major difference relative to control runs, where there are no massive clumps that can cause ejections.

Without BH accretion and feedback the outcome is dependent on the orbital eccentricity. This is consistent with result on nuclear discs, which also found the eccentricity of the orbit to be very important \citep[e.g.][]{Fiacconi2013,DelValle2014}. Looking at Figure~\ref{fig:separation_massive_f0} in more detail, we can see that, for the case of the circular orbit (top panel), the BH close-pairing time is fast, i.e. 146~Myr. Remarkably, the decay is faster in this case relative to the control run, at odds with the general trend towards a delayed orbital decay. We explain this by analogy with the results of migrating gas giant planets in massive self-gravitating discs, where it is found that the non-linear disc torques, mainly originating from the co-orbital region, increase steeply with increasing gas density \citep{Malik_et_al_2015}. While we postpone a detailed torque analysis to a forthcoming paper, the analogy should hold because, as we have already explained, on circular orbits disc torques rather than dynamical friction drive the orbital decay. Finally, we note that in this case the decay is so fast that the probability of scattering on to massive clumps, that are still rare at this early stage, is very low, allowing to describe the decay as if clumps were not present (the secondary BH encounters lower-mass clumps that are seen to be disrupted by its tidal force on its way to the centre). In addition, on circular orbits the velocity difference between clumps and BHs will be statistically smaller, hence the torques exerted by clumps will be smaller.

When we add BH accretion and feedback, the lower density of the gas weakens disc torques, slowing down the decay. As a result, the secondary BH orbits for a longer time in the fragmenting disc, becoming more susceptible to perturbations by massive clumps. This explains why the character of the decay changes dramatically, leading to a clear delay. The peak at $\sim 500$~Myr in the second panel of Figure~\ref{fig:separation_massive_f0}, for example, is due to an interaction with an individual clump. In the third panel of Figure~\ref{fig:separation_massive_f0} we reduce the BH feedback efficiency $\epsilon_{\rm f}$ from 0.001 to 0.0001. We would  expect that, as in the control case, having a lower $\epsilon_{\rm f}$ would yield a shorter orbital decay time-scale, but this does not occur because the perturbations by massive clumps dominate the orbital dynamics in this case, scattering the secondary BH into the low-density envelope around the disc plane.

The effect of varying eccentricity does not reveal an overall systematic trend in clumpy runs when BH accretion and feedback are included. This is because the orbital dynamics loses memory of the initial conditions very quickly due to the repeated perturbations by massive clumps and strong spiral waves. Indeed, as it is evident from Figure~\ref{fig:separation_clumpy}, the eccentricity evolves stochastically in each single run independently of the initial value. The tendency is towards maintaining an excitation of the eccentricity until the end of the simulation.

%
%%%%%%%%%%%%%%%%%%%%%%%%%%%%%
%

\subsection{Effect of a clumpy ISM on BH mass growth}

%
%%%%%%%%%%%%%%%%%%%%%%%%%%%%%
%

\begin{figure*}
\flushleft
\includegraphics[width=0.49\textwidth]{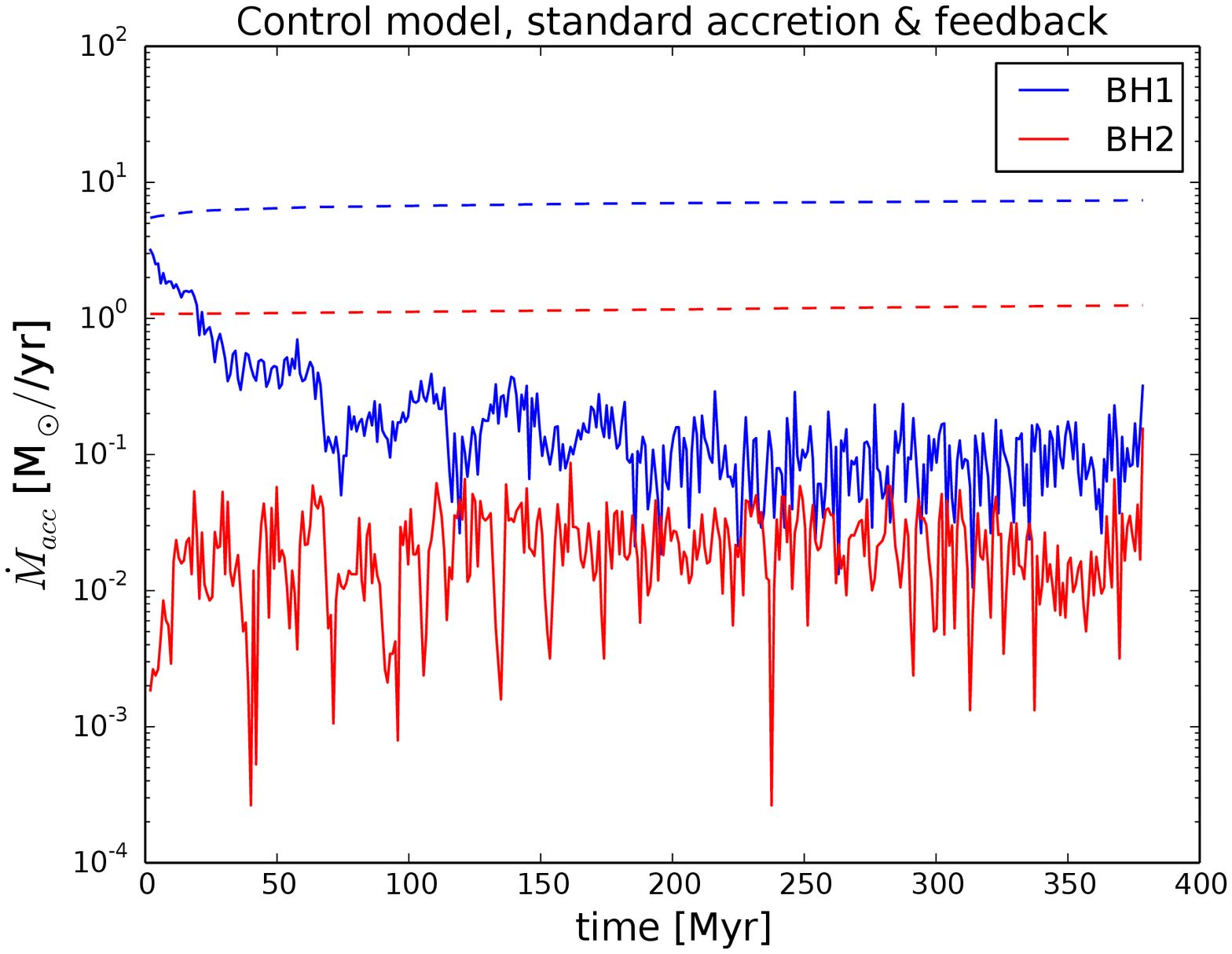}
\includegraphics[width=0.49\textwidth]{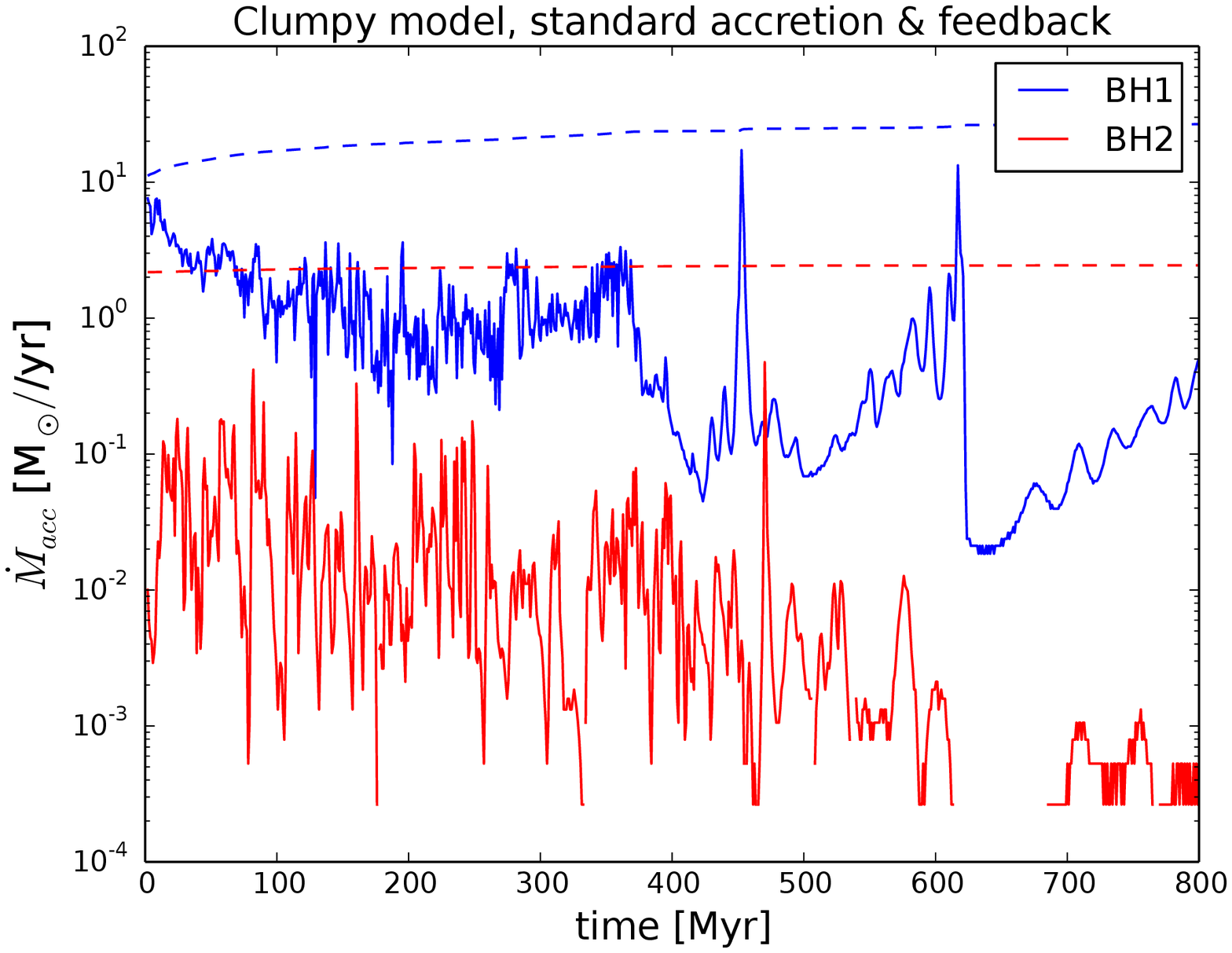}
\caption{Time evolution of the mass accretion rate (solid lines) for the primary (blue) and secondary (red) BH in the control model (left panel) and clumpy model (right panel) with standard BH accretion and feedback ($\alpha = 1$, $\epsilon_{\rm f} = 0.001$, $\epsilon_{\rm r} = 0.1$) and zero eccentricity. BH Eddington accretion rates are also shown (dashed lines). The BH accretion rate in the clumpy model is a factor 10 higher than in the control case and shows peaks due to accretion of clumps, see the text for more details. The final BH masses are $3.3 \times 10^8$~M$_{\odot}$ and $5.6 \times 10^7$~M$_{\odot}$ (control model), $1.2 \times 10^9$~M$_{\odot}$ and $\sim 10^8$~M$_{\odot}$ (clumpy model).}
\label{fig:mass_accretion_compared}
\end{figure*}

By comparing our clumpy runs with their corresponding control runs, which produce very few, short-lived clumps, we  can single out the effect of a clumpy ISM on BH mass growth. Consider the most simple case, with standard BH accretion and feedback, and zero eccentricity. In the control run, when the close BH pair forms, the primary BH grows up to $3.3 \times 10^8$~M$_{\odot}$, while the secondary up to $5.6 \times 10^7$~M$_{\odot}$, less than a factor 1.5 from the beginning of the simulation. In the corresponding clumpy run, on the other hand, the mass of the primary BH at the end of the simulation is $1.2 \times 10^9$~M$_{\odot}$ (the mass of the secondary is still low, $\sim 10^8 $~M$_{\odot}$), more than a factor 2 larger than the initial mass, already reached at $\sim 378$~Myr, the time at which a close pair forms in the control run.

By inspection of  Figure~\ref{fig:mass_accretion_compared} (which is a representative comparison of all our simulations), we can see that the mass accretion in the control run is almost a factor of 10 lower than that in the clumpy run. There are at least two reasons behind the different mass accretion rates. First of all, the primary BH in the clumpy model is initialised to be about a factor of 2 more massive than that in the control run. This implies that, just based on the Bondi accretion dependence on mass, we expect $\dot{M}_{\rm acc}$ to be 4 times higher in the clumpy run. The average gas density in the more massive disc with higher gas fraction in the clumpy run is also a factor of $\sim 2$ higher than in the control run, which yields an increase of another factor of $\sim 2$ in the Bondi rate. Therefore, simply the higher BH mass and gas density alone account for nearly all the difference seen in the BH accretion rate between the two runs. Considering also that the clumpy disc is more strongly self-gravitating and therefore must support a more vigorous gas inflow rate through spiral-wave driven torques, there appears to be very little remaining room for a direct role of clumps in the higher accretion rate. A caveat is, of course, that the concurrent role of BH feedback will alter the simple behaviour of the mass accretion rate underscored by the dependencies in the Bondi rate.

The presence of clumps is however discernible from the more episodic character of mass accretion relative to the control run, with  peaks seemingly associated with the accretion of the two most massive clumps ($M_{\rm cl-gas} \sim 4 \times 10^8$ and $1.4 \times 10^8$~M$_{\odot}$), which migrate to the galaxy centre and are accreted by the primary BH.

As one can see, our BHs never accrete at Eddington rate. In a previous work, \citet{GaborBournaud2013} found that the mass accretion rates for the central BH are generally lower than in our case: indeed in their simulations the BH accretion rate is generally $10^{-5}$~M$_{\odot}$~yr$^{-1}$, whereas we find $\sim 10^{-1}$~M$_{\odot}$~yr$^{-1}$ in the control case and $\sim 1$~M$_{\odot}$~yr$^{-1}$ in the clumpy case. The substantial difference is that they have peaks that reach the Eddington rate, whereas we do not. We can compare our control and clumpy runs with standard BH accretion and feedback to their models M16f10 and M4f50, respectively. This behaviour has likely two reasons: first of all, in \citet{GaborBournaud2013}, the aim was studying the BH growth in a clumpy medium, so they have only one central BH, into which all the produced clumps can accrete, while we have also the second BH that perturbs the medium and can accrete clumps as well. The massive perturber is very important in the control case, where gravity is not dominant like in the clumpy case, and indeed our galaxy has spiral arms also in the central part and a higher inflow, whereas the model M16f10 of \citet{GaborBournaud2013} does not: the gas density map is quite smooth (see their Figure~1, bottom-left panel), except for some clumps in the outer part, maybe due a different relaxation phase of the disc or to the short simulation time. Moreover, \citet{GaborBournaud2013} have BHs ten time less massive than in our models (and therefore a lower Eddington mass accretion rate), which could explain the observed peaks at Eddington rates in their mass accretion rates.

%
%%%%%%%%%%%%%%%%%%%%%%%%%%%%%
%

\subsection{Limits of the isolated simulations; the ``asymptotic'' decay time-scale}

%
%%%%%%%%%%%%%%%%%%%%%%%%%%%%%
%

%

In our initial conditions, the disc scale height is $0.05\, R_{\rm d}$, where $R_{\rm d} \sim 2$~kpc. During the simulations, the vertical disc extent remains below 1~kpc, even during the vigorous fragmentation phase of the clumpy disc model (see Figure~\ref{fig:low_density_region}, edge-on views). This implies that, when the secondary BH is ejected out of the disc plane, the drag by dynamical friction drops dramatically, and the orbital decay is correspondingly suppressed. As a result, a close BH pair never forms when the secondary BH is ejected out of the disc: this happens in some of the clumpy-disc runs. However, by construction our galaxy models lack an extended dense spheroidal component, such as a stellar bulge, whose presence would increase the background stellar density above and below the disc plane. It is thus important to consider the effect of such a component on the orbital decay of the BHs that are ejected out of the disc. In passing, we note that a small exponential pseudo-bulge forms as a result of gas inflows during the relaxation phase of our initial conditions (with mass $\sim 4 \times 10^9$~M$_{\odot}$) and later grows even more by further inflows and inward clump migration during the fragmentation phase, as discussed in \citet{Tamburello2015}. However, such a bulge is flat and disky, extending less than 200~pc from the midplane, and thus yields no extra contribution to the stellar density beyond the disc midplane.

For this reason, in this last section, we investigate the effect of adding an extended spherical stellar bulge to the calculation of the orbital decay time-scale in the cases when, during the simulations, the secondary BH sometimes moves out of the disc into a background of stars, losing orbital angular momentum due to dynamical friction (\citealt{SouzaLima2016}, in prep.). The \citet{Chandrasekhar_1943} dynamical friction (DF) formula gives us an equation for the force to which the BH is subject:

\begin{equation}
\mathbf{F}_{\rm DF} = - 16 \pi^2 G^2 M^2_{\rm BH}m_{\rm a} \ln{\Lambda} \left[ \int_0^V v_{\rm a}^2 f(v_{\rm a}) dv_{\rm a} \right] \frac{\mathbf{v}}{v^3},
\label{eq:int}
\end{equation}

\noindent where $\mathbf{v}$ is the velocity of the BH relative to the background, $m_{\rm a}$ is the individual mass of the particles in the background, $v_{\rm a}$ their velocity, $f(v_{\rm a})dv$ is the number of particles with velocity between $v$ and $v+dv$, and $\ln{\Lambda}$ is the Coulomb logarithm.

Now, consider a BH subject to dynamical friction exerted by a stellar bulge described by a \citet{Hernquist_1990} density profile, described by

\begin{equation}
\rho_{\rm H} = \frac{M_{\rm b}}{2\pi} \frac{a}{r}\frac{1}{(r+a)^3},
\end{equation}

\noindent where $M_{\rm b}$ is the total bulge mass and $a$ is its scale radius. If we also consider a Maxwellian distribution function for the velocity and assume $v$ to be large enough so that we can estimate the integral term in Equation~\eqref{eq:int} up to infinity (within a factor $\xi$), the integral converges to the number density of stars, $n_*$, divided by $4\pi$, and the expression for the force becomes

\begin{equation}
\mathbf{F}_{\rm DF} = - 2 G^2 M^2_{\rm BH} \ln{\Lambda} \xi M_{\rm b} \frac{a}{r}\frac{1}{(r+a)^3} \frac{\mathbf{v}}{v^3},
\label{eq:Fdf2}
\end{equation}

\noindent after having used $m_{\rm a} n_* = \rho_{\rm H}$. For simplicity, we consider the BH on a circular orbit, so that $v = v_{\rm c}$, where $v_{\rm c}$ is the circular velocity calculated in the Hernquist model as

\begin{equation}
v_{\rm c} = \frac{\sqrt{GM_{\rm b}r}}{r+a}.
\label{eq:vcirc}
\end{equation}

Because the force is perpendicular to the radial direction and produces a torque, we can substitute Equation~\eqref{eq:vcirc} into Equation~\eqref{eq:Fdf2} to obtain an expression for the torque:

\begin{equation}\label{eq:torque}
r F_{\rm DF} = - 2 G M^2_{\rm BH} \ln{\Lambda} \xi \frac{a}{r}\frac{1}{(r+a)}.
\end{equation}

The torque applied on the BH corresponds to the time variation of its angular momentum, $L$:

\begin{eqnarray}\label{eq:ang_mom}
\frac{dL}{dt} & = &\frac{d}{dt}(M_{\rm BH}rv) \\ & = & M_{\rm BH} \sqrt{GM_{\rm b}} \frac{\dot{r} r^{1/2}}{(r+a)^2} \frac{r+3a}{2}. \nonumber
\end{eqnarray}

We can then equate the two equations above, $dL/dt = rF_{\rm DF}$, to obtain an equation for $\dot{r}$ and then, integrating, we find an estimate for the dynamical friction time-scale defined as the time required for the orbital angular momentum to go to zero:

\begin{eqnarray}\label{eq:time}
t_{\rm DF} &=& 30 \; \xi^{-1} {\rm Myr} \left( \frac{\ln{\Lambda}}{5}\right) \left( \frac{M_{\rm b}}{10^9\, {\rm M}_{\odot}}\right)^{1/2} \\ &\cdot& \left( \frac{M_{\rm BH}}{5 \times 10^7\, {\rm M}_{\odot}}\right)^{-1} \left( \frac{a}{200\; {\rm pc}}\right)^{-1} \left( \frac{r_{\rm init}}{1\; {\rm kpc}}\right)^{5/2}. \nonumber
\end{eqnarray}

With this equation, where $M_{\rm BH}$ is the secondary BH's initial mass (we used the initial mass, because we have seen that the secondary BH does not grow much during the simulations) and $r_{\rm init}$ is the distance from the centre the BH finds itself at, we try to estimate the orbital decay time-scale for the clumpy model, using a range of values for $a$ and $M_{\rm b}$ (see Table~\ref{table3}). Notice that, in order to obtain Equation~\eqref{eq:time}, we used the approximation $r_{\rm init} \gg a$, which fits our typical values, but in the opposite case, $r_{\rm init} \ll a$, the equation is still valid simply multiplying the left-hand side by a factor of 3. In both approximations, the orbital decay time-scale decreases when increasing the bulge scale radius $a$ or decreasing the bulge total mass $M_{\rm b}$. When $r_{\rm init} \ll a$, the dynamical friction force is independent of $a$, but the angular momentum of the BH decreases with $a$ and so the time-scale. In the case $r_{\rm init} \gg a$, instead, the angular momentum is independent of $a$, whereas the dynamical friction force increases with $a$, so the time-scale decreases. Analogously, the dynamical friction force is independent of $M_{\rm b}$ and the angular momentum increases with the bulge mass, so the time-scale increases. Table~\ref{table3} shows that the dynamical friction time-scales due to the bulge are shorter than 0.1~Gyr in general, assuming, as we have done, that the secondary BH starts at a radius of 1~kpc from the centre, this being a typical distance after ejection (see Figure~\ref{fig:low_density_region}). 

Therefore, a bulge would indeed shorten the time the secondary BH finds itself outside the disc. The subsequent  evolution of the orbit depends on the relative timing of formation of clumps and bulge. If a bulge forms before the clumps and the secondary BH is already outside the disc (e.g. as the final outcome of a merger), the BH would then quickly fall towards the centre. However, if clumps are also present (and, as in our simulations, are the cause of ejection), the secondary BH could interact again with them, which can either (i) expel it again from the disc, or (ii) simply launch the BH onto a much wider orbit at the periphery of the disc, via gravitational slingshots (as in, e.g., our clumpy run with zero eccentricity and standard BH accretion and feedback).

%
%%%%%%%%%%%%%%%%%%%%%%%%%%%%%
%

\section{Summary and discussion}

%
%%%%%%%%%%%%%%%%%%%%%%%%%%%%%
%

\begin{table}
\vspace {1 mm}
\centering
\setlength{\extrarowheight}{0.1cm}
\begin{tabular}{x{1.5cm}|x{1.5cm}|x{1.5cm}|x{1.5cm}|x{1.5cm}}
\diag{.1em}{1.5cm}{$M_{\rm b}$}{$a$}  & 300 pc & 400 pc \\
\hline
$10^9$~M$_{\odot}$ 		 &  10.15  & 7.61 \B \\
$5 \times 10^9$~M$_{\odot}$  & 22.7   &  17.03 \B  \\
\end{tabular}
\caption{The table shows the estimated dynamical friction time-scale [in units of $(r_{\rm init}/1~{\rm kpc})^{5/2}$~Myr], using Equation~\eqref{eq:time} and assuming $\ln{\Lambda} = 5$ and $\xi = 1$, for the clumpy model, where the secondary BH mass is $9.8 \times 10^7$ $M_{\odot}$, for $a$ in the range $300-400$~pc and $M_{\rm b}$ in the range $10^9 - 5 \times 10^9$~M$_{\odot}$.}
\label{table3}
\end{table}

The role of gravitational interactions with clumps as well as that of BH accretion and feedback are both very important for the orbital decay. Looking at the results of the various runs in Table~\ref{table2} and recalling that the control model fragments only moderately and becomes stable after $150$~Myr, while the clumpy model fragments vigorously for longer than 500~Myr, we can distinguish 4 cases:

\begin{enumerate}
	\item control model - without BH accretion and feedback: a close BH pair, which eventually will bind into a binary, always forms;
	\item control model - with BH accretion and feedback: a close BH pair always forms, but with a delay of about a factor of 2;
	\item clumpy model - with strong interactions with clumps: a close BH pair never forms;
	\item clumpy model - without strong interactions with clumps: a close BH pair always forms, with the timing depending on the presence of BH accretion and feedback.

\end{enumerate}

In the third case we can have either ejections (e.g. runs Clumpy-f0-highAcc-stdFB and Clumpy-f1-noAcc; see bottom panels of Figure~\ref{fig:low_density_region}) or slingshots (e.g. runs Clumpy-f0-stdAcc-stdFB and Clumpy-f2-noAcc) of the secondary BH due to interactions with massive clumps. In case of ejection, the secondary BH finds itself in a much lower density region, where dynamical friction becomes very weak and disc torques are absent. 
In the fourth case the role of BH accretion and feedback is important and causes a delay in the BH pair formation. In runs Clumpy-f0-noAcc and Clumpy-f02-noAcc, for example, which have no BH accretion, a close BH pair forms earlier than in runs Clumpy-f02-stdAcc-stdFB, Clumpy-f1-stdAcc-stdFB, and Clumpy-f2-stdAcc-stdFB, which have BH accretion and feedback. This is due to the weaker drag in the more rarefied gaseous background heated by BH feedback.

The actual values of the orbital decay time-scales in the clumpy runs should be taken with caution since our simulations model isolated galaxies while at high redshift interactions, mergers and rapid, prominent gas accretion will always take place. Moreover, the calculations of the last section show that the effect of a (dense) bulge could be crucial in re-starting the decay and bringing the secondary BH back into the disc midplane in less than 1~Gyr. While many of the clumpy high-redshift star-forming galaxies do not exhibit clear evidence that they do possess already a bulge, migration of massive clumps to the centre has been advocated often as a mechanism to grow such a bulge \citep[e.g.][]{Bournaud2016}. Compact ``red nuggets'' are observed  at similar redshifts as clumpy, star forming galaxies. These red nuggets could be bulges that have already begun to quench \citep{Barro2013}. An evolutionary connection between red nuggets and clumpy star-forming galaxies has been attempted, and probably requires a combined role of mergers and disc fragmentation \citep{Dekel2014}. If such connection exists and red nuggets are related to bulges, then red nuggets could be the result of rapid quenching of star-forming clumpy galaxies, which would suggest a short duration of the clumpy phase, possibly of order a Gyr. Likewise, in the last section we have shown that once the bulge is present, the secondary BH returns to the plane in less than 0.1~Gyr. Therefore, once the host galaxy enters a red-nugget phase, it is conceivable that the orbital decay will promptly restart and lead to the formation of a BH binary in less than 1~Gyr. The stalling phase would thus last at most up to the time of bulge formation, which appears to happen not more than 1~Gyr after the epoch of clumpy high-redshift galaxies.

Furthermore, clumps always excite the eccentricity of the BH pair, which is often not damped until the end of the simulation. This is interesting since it suggests that the orbit of the secondary BH, if it returns to the midplane and restarts decay, will have a high eccentricity no matter what the orbit was soon after the galaxy merger. Thus, in this case, gas does not lead to circularisation, but rather the opposite. If the role of stellar encounters becomes important at smaller binary separations, these will keep the eccentricity high, possibly delivering a tight eccentric binary to the final GW emission stage, which then will occur on a very fast track.

The results of this work, when combined with the results obtained on the scale of BH pairs at smaller separations in circumnuclear discs \citep[e.g.][]{Fiacconi2013}, as well as with those of multi-scale gas-rich  mergers probing sub-pc scale separations in both smooth and clumpy ISM conditions \citep{Chapon2013,Roskar2015} suggest that the orbital decay of BH pairs in gaseous backgrounds occurs in a variety of regimes, some of which lead to a slow orbital decay that is not more efficient than that found in modern calculations of the orbital decay and hardening of BH binaries embedded in stellar distributions where partial loss cone refilling occurs owing to centrophilic stellar orbits in triaxial potentials \citep{Khan_et_al_2011,Khan_et_al_2012,Sesana_Khan_2015}. This is contrary to the conventional notion that gaseous backgrounds speed-up the orbital decay process significantly already from the BH binary formation stage and avoid the last parsec problem altogether \citep{Mayer2013}. The stellar dynamical simulations find BH merger time-scales of order a few Gyr for stellar hosts akin to present-day galaxies, where the time-scale is estimated by extrapolating to the separation at which the GW emission phase starts and leads to prompt coalescence \citep{Sesana_Khan_2015}. Estimating an overall coalescence time-scale for the BHs in our simulations is not trivial since, as we elaborated, the host may enter a quenched, red-nugget phase in which the binary may finally form and then harden due to the interaction with a stellar background, exactly because in a quenched system the contribution of stars to the mass will become dominant relative to that of the ISM. However, existing calculations of binary hardening in stellar background assume stellar potential akin to those of present-day galaxies, hence not directly applicable to our situation. This will have to be explored with future multi-scale calculations of binary BHs embedded in realistic models of high-redshift galaxies. For now, we can only assert that the first stage of the orbital decay, namely the formation of the binary BH, will occur on a time-scale of a few times $10^8$~yr to 1~Gyr in clumpy galaxies, The overall BH orbital decay time up to final coalescence can only be longer than this. These are long time-scales at an epoch, such as $z \sim 2$, when the lookback time is only $\sim 3$~Gyr.

We argue that, between $z \sim 3$ and $\sim 1$, namely up to the epoch at which the fraction of clumpy galaxies appears to drop \citep{FoersterSchreiber2011, Guo2015}, there is a phase in which the effects described in this paper on the cosmic population of massive BH pairs are important. In such a case the probability of having wandering BHs is maximal, as they are waiting to restart decay after the spheroid has been established. If such a phase exists, since separations between BHs can become of a few kpc, the probability of detecting bright dual AGN (the BHs in massive clumpy galaxies will be at least as massive as we assumed here, hence powerful radiation sources) will be high. At the same time, such phase is the least favourable epoch for detecting BH mergers with GW experiments. While eLISA is primarily sensitive to lighter BHs, which would be hosted in galaxies more similar to our control run, thus in principle not subject to delayed orbital decay in clumpy discs, it has been recently pointed out that Pulsar Timing Arrays (PTAs) would be sensitive to BH mergers in the mass range of those considered in this work over a wide range of redshifts \citep[$z < 2$,][]{Sesana09}. Additionally, \citet{Rosado16} claim that detection by PTAs should be possible at arbitrarily high redshift, assuming BH binaries of mass $> 10^{10}$~M$_{\odot}$. These masses are larger than those consider in this work. However, such binaries would likely be found in comparably if not more massive galaxy disks compared to those modelled here. For even more massive galaxies the clumpy phase should still be relevant, perhaps occurring even earlier.

Finally, we stress the fact that BH feedback alone leads to a delay in the BH binary formation even in relatively smooth galaxy hosts as those of the control runs. While the measured delay is moderate, about a factor of 2, it remains to be assessed how the quantitative result will change for galaxy hosts and BHs more in line with the typical targets that eLISA will probe, namely BHs with mass below $10^7$~M$_{\odot}$, which would then likely reside in galaxies about a factor of $5-10$ less massive than the models adopted in the control run. Since not only the BH masses and galaxy masses would be different, but also the effect of feedback will change (if anything because the accretion rate on lighter BH will be different and the ambient density and pressure will be also different), new simulations targeted to the problem will be needed.

%
%%%%%%%%%%%%%%%%%%%%%%%%%%%%%
%

\section*{Acknowledgements}
This work is supported by the STARFORM Sinergia Project funded by the Swiss National Science Foundation. The authors thank the anonimous referee for helpful commenst and suggestions. PRC acknowledges support by the Tomalla Foundation and thanks Massimo Dotti for helpful suggestions. LM thanks Alberto Sesana for useful discussions about the implications for the detection of BH binaries with GW experiments. VT thanks Fr\'ed\'eric Bournaud, Davide Fiacconi and Rafael Souza Lima for helpful discussions.

%
%%%%%%%%%%%%%%%%%%%%%%%%%%%%%
%

\scalefont{0.94}
\setlength{\bibhang}{1.6em}
\setlength\labelwidth{0.0em}
\bibliography{Paper2}
\bibliographystyle{mnras}
\normalsize

%
%%%%%%%%%%%%%%%%%%%%%%%%%%%%%
%

\appendix

%
%%%%%%%%%%%%%%%%%%%%%%%%%%%%%
%

\section{Deriving the dynamical friction time-scale}

%
%%%%%%%%%%%%%%%%%%%%%%%%%%%%%
%

In this section, we show all the necessary steps to obtain Equation~\eqref{eq:time}, starting by equating Equations~\eqref{eq:ang_mom} and \eqref{eq:torque}, $dL/dt = rF_{\rm DF}$, and solving for $\dot{r}$:

\begin{eqnarray}
\dot{r} & = & - \frac{1}{\left(M \sqrt{GM_{\rm b}}\right)} \frac{(r+a)^2}{r^{1/2}} \frac{2}{r+3a} 2GM^2_{\rm BH} \ln \Lambda \xi \frac{a}{r} \frac{1}{r+a} \nonumber \\
           & = & - 4 \xi \ln \Lambda \sqrt{ \frac{G}{M_{\rm b}} } M_{\rm BH} a \frac{r+a}{r^{3/2} (r+3a)} \nonumber \\
           & = & - 4 \xi \ln \Lambda \sqrt{a^2 G M_{\rm b}} \frac{M_{\rm BH}}{M_{\rm b}} \frac{r+a}{r^{3/2} (r+3a)}. \nonumber \\
\end{eqnarray}

If we now solve for $t$, we can find the decay time-scale, $t_{\rm DF}$, integrating from an initial radius $r_{\rm init}$ to a final radius $r_{\rm f} = 0$.

\begin{eqnarray}\label{eq:integral}
t_{\rm DF} & = & - \frac{1}{4 \xi \ln \Lambda} \frac{1}{\sqrt{a^2 G M_{\rm b}}} \frac{M_{\rm b}}{M_{\rm BH}} \int_{r_{\rm init}}^{r_{\rm f}} \frac{r^{3/2} (r+3a)}{r+a} dr \nonumber \\
         & = & \frac{1}{4 \xi \ln \Lambda} \frac{1}{\sqrt{a^2 G M_{\rm b}}} \frac{M_{\rm b}}{M_{\rm BH}} \int_{\chi_{\rm f}}^{\chi_{\rm init}} \frac{a^{5/2} \chi^{3/2} (\chi + 3) } {a (\chi + 1)} a d\chi \nonumber \\
         & = & \frac{1}{4 \xi \ln \Lambda} \frac{1}{\sqrt{a^2 G M_{\rm b}}} \frac{M_{\rm b}}{M_{\rm BH}} a^{5/2} \int_{\chi_{\rm f}}^{\chi_{\rm init}} \frac{\chi^{3/2} (\chi + 3) } {\chi + 1} d\chi \nonumber \\
         & = & \frac{1}{4 \xi \ln \Lambda} \sqrt{ \frac{a^3}{G M_{\rm b}}} \frac{M_{\rm b}}{M_{\rm BH}} \int_{\chi_{\rm f}}^{\chi_{\rm init}} \frac{\chi^{3/2} (\chi + 3) } {\chi + 1} d\chi, \nonumber \\
\end{eqnarray}

\noindent where $\chi = r/a$. The integral in Equation~\eqref{eq:integral} has two limiting solutions: in our case $\chi \gg 1$, so the integral can be approximated to $2/5 \chi^{5/2}$, while in the case $\chi \ll 1$ it becomes $6/5 \chi^{5/2}$. Substituting in Equation~\eqref{eq:integral} we find:

\begin{equation}
t_{\rm DF} = \frac{1}{10 \xi \ln{\Lambda}} \sqrt{\frac{a^3}{GM_{\rm b}}} \frac{M_{\rm b}}{M_{\rm BH}} \left( \frac{r_{\rm init}}{a}\right)^{5/2}.
\end{equation}

\end{document}